\RequirePackage{fix-cm} 
\documentclass[smallextended]{svjour3} 
\smartqed  

\makeatletter
\def\cl@chapter{\@elt {theorem}}
\makeatother

\usepackage{cleveref} 
\journalname{}
\usepackage{lineno,hyperref}
\usepackage{mathalfa}
\newcommand{\tool}{\textsc{MKLDroid}}
\newcommand{\soa}{state-of-the-art}
\usepackage{url}
\usepackage{inconsolata}
\usepackage{mdframed}
\mdfdefinestyle{MyFrame}{%
	linecolor=black,
	outerlinewidth=2pt,
	roundcorner=20pt,
	innertopmargin=\baselineskip,
	innerbottommargin=\baselineskip,
	innerrightmargin=20pt,
	innerleftmargin=20pt}
\usepackage{multirow}
\usepackage[table,xcdraw]{xcolor}
\usepackage[utf8]{inputenc}
\usepackage{cleveref}
\crefname{section}{§}{§§}
\Crefname{section}{§}{§§}
\usepackage{float}
\usepackage{graphicx}
\usepackage{hyperref}
\usepackage{color}
\usepackage[space]{cite}
\hypersetup{
	colorlinks   = true,
	citecolor    = blue,
	urlcolor     = blue,
	linkcolor	 = blue
}
\makeatletter
\DeclareUrlCommand\ULurl@@{%
	\def\UrlLeft{\uline\bgroup}%
	\def\UrlRight{\egroup}}
\def\ULurl@#1{\hyper@linkurl{\ULurl@@{#1}}{#1}}
\DeclareRobustCommand*\ULurl{\hyper@normalise\ULurl@}
\makeatother

\usepackage{array}
\newcolumntype{L}[1]{>{\raggedright\let\newline\\\arraybackslash\hspace{0pt}}m{#1}}
\newcolumntype{C}[1]{>{\centering\let\newline\\\arraybackslash\hspace{0pt}}m{#1}}
\newcolumntype{R}[1]{>{\raggedleft\let\newline\\\arraybackslash\hspace{0pt}}m{#1}}

\usepackage{caption}
\usepackage{subcaption}
\usepackage{amsfonts}
\usepackage{fixltx2e}
\usepackage{bm}
\usepackage{amsmath}
\usepackage{mathrsfs}
\usepackage{bbm}
\usepackage{dsfont}
\usepackage{yfonts}
\usepackage{graphicx}
\usepackage{subcaption}
\usepackage{amssymb}
\usepackage{tikz}
\usetikzlibrary{shapes.geometric, arrows}
\tikzstyle{ADG} = [ellipse, minimum width=1cm, minimum height=.75cm,text centered, draw=black, fill=pink!30]
\tikzstyle{PDG} = [ellipse, minimum width=1cm, minimum height=.75cm,text centered, draw=black, fill=yellow!20]
\tikzstyle{SSCFP} = [ellipse, minimum width=1cm, minimum height=.75cm,text centered, draw=black, fill=green!10]
\tikzstyle{Ins} = [ellipse, minimum width=1cm, minimum height=.75cm,text centered, draw=black, fill=brown!10]
\tikzstyle{Signs} = [ellipse, minimum width=1cm, minimum height=.75cm,text centered, draw=black, fill=orange!10]

\usepackage{pifont}
\usepackage[ruled,vlined,linesnumbered]{algorithm2e}

\usepackage{verbatim}
\usepackage{algpseudocode}
\usepackage{amsmath}
\usepackage{enumitem}
\setlist[enumerate]{leftmargin=*}
\usepackage[space]{cite}
\usepackage{footmisc}
\usepackage{footnote}
\makeatletter
\newtoks\therules% Contains rules
\therules={}% Start with empty token list
\def\appendto#1#2{\expandafter#1\expandafter{\the#1#2}}% Append to token list
\def\gobblefirst#1{% Remove (first) from token list
	#1\expandafter\expandafter\expandafter{\expandafter\@gobble\the#1}}%
\def\LState{\State\unskip\the\therules}% New line-state
\def\printindent{\unskip\the\therules}%

\usepackage{xcolor}
\usepackage[linesnumbered,ruled,vlined]{algorithm2e}

\SetCommentSty{mycommfont}

\usepackage{setspace}
\usepackage{esvect}

\begin{document}
	
\sloppy

	\title{A Multi-view Context-aware Approach to Android Malware Detection and Malicious Code Localization
	}

	\author{Annamalai Narayanan         \and
	Mahinthan Chandramohan  \and
	Lihui Chen \and
	Yang Liu
	}
	
	\institute{
		annamala002@e.ntu.edu.sg, \{mahinthan,elhchen,yangliu\}@ntu.edu.sg \\
		Nanyang Technological University\\
		Singapore
	}
	
	\maketitle

%\vspace{-5mm}
\begin{abstract}
	
	\color{black}{Existing Android malware detection approaches use a variety of features such as security-sensitive APIs, system calls, control-flow structures and information flows in conjunction with Machine Learning classifiers to achieve accurate detection. Each of these feature sets provides a unique semantic \textit{perspective} (or \textit{view}) of apps' behaviors with inherent strengths and limitations. Meaning, some views are more amenable to detect certain attacks but may not be suitable to characterize several other attacks. 
		%{\color{red} For instance, information flow features are amenable for detecting '\textit{privacy leaks}' but incapable of revealing '\textit{privilege escalations}'}. 
		Most of the existing malware detection approaches use only one (or a {\color{black}selected few}) of the aforementioned feature sets which prevents them from detecting a vast majority of attacks. Addressing this limitation, we propose \tool, a unified framework that systematically integrates multiple views of apps for performing comprehensive malware detection and malicious code localization.
		%, namely, API and permission dependencies, instruction sequences, control flow structures and information source and sinks. 
{\color{black}The rationale is that, while a malware app can disguise itself in some views, disguising in every view while maintaining malicious intent will be much harder}. 
		
		\tool{} uses a graph kernel to capture structural and contextual information from apps' dependency graphs and identify malice code patterns in each view. Subsequently, it employs Multiple Kernel Learning (MKL) to find a weighted combination of the views which yields the best detection accuracy. Besides multi-view learning, \tool’s unique and salient trait is its ability to locate fine-grained malice code portions in dependency graphs (e.g., methods/classes). Malicious code localization caters several important applications such as supporting human analysts studying malware behaviors, engineering malware signatures, and other counter-measures. 
		%Malicious code localization becomes particularly useful in the case of \textit{repackaged malware}, as a predominant portion of their code remains benign and only a small portion is attack related. 
		Through our large-scale experiments on several datasets (incl. wild apps), we demonstrate that \tool{} outperforms three \soa{} techniques consistently, in terms of accuracy while maintaining comparable efficiency. In our malicious code localization experiments on a dataset of repackaged malware, \tool{} was able to identify all the malice classes with 94\% average recall. 	
		Our work opens up two new avenues in malware research: (i) enables the research community to elegantly look at Android malware behaviors in multiple perspectives simultaneously, and (ii) performing precise and scalable malicious code localization.}
	
	\keywords{Android Malware Detection, Graph Kernels, Multiple Kernel Learning, Malicious Code Localization}
	
\end{abstract}

\section{Introduction}
\label{sec:intro}

Over the past few years, proliferation of malware for mobile platforms such as Android has been severe. For instance, Kaspersky reports \cite{ks} discovering nearly 4.1 million new malware in 2016 which is a 17\% increase over 2015.  A major reason for such tremendous volume and growth rate is the production of \textit{repackaged malware variants}. Typically, these variants are produced through repackaging popular legitimate applications (apps) with similar malicious code. The Android app packaging and distribution model offers painless and straightforward opportunities for attackers to piggyback (i.e., inject) their malice code on benign apps which could then be spread through several third-party markets. Many empirical studies \cite{massvet,drdroid,grafting,hookranker} manifest that an overwhelming majority of Android malware are nothing but repackaged versions of benign apps. 
%In fact, more than 80\% of samples in a well-known primeval Android malware dataset named Genome \cite{Genome} are repackaged apps. 
%Therefore, peculiarly, in these mobile malware, a predominant portion of code remains benign and only a small portion is attack related. 
The sheer volume, growth rate and evolution of repackaged and other types of malware highlight an imperative need for developing effective and scalable automated malware detection techniques \cite{grafting,hookranker,drdroid}.

%A recent work named \textsc{DrDroid} \cite{drdroid} attempts to localize malice code by splitting an app's FCG into multiple regions called as DRegions and predicts whether each of them is malicious or benign. However, on many occasions, this method finds only one DRegion in apps, thus labeling all the code in an app as either malicious or benign. 
%In fact, out 5,600 apps in \textsc{Drebin} \cite{Drebin}, a well-known benchmark dataset, \textsc{DrDroid} marked the entire code as malicious in 3,757 apps. Also, this approach could not rank portions of apps such as methods or classes based on this severity or degree of malicousness. 

To perform automated detection, recent approaches  from both academia and industry increasingly resort to program analysis and machine learning (ML) techniques. Typically, the detection process involves extracting semantic features from suitable representations of programs (e.g., assembly code, call graphs (CGs), etc.) and identifying malicious code/behaviour patterns using ML classifiers \cite{Drebin,droidscribe,droidseive,CSBD,Adagio,DroidMiner,AppContext,DroidSift,MLMalDetect}. Notably, higher level semantic representations such as CGs, control- and data-flow graphs mostly stay similar even when the code is considerably altered \cite{DroidMiner,DroidSift,Adagio,CSBD,MLMalDetect} (we use a common term 'Program Representation Graph' (PRG) to refer to any of the aforementioned graphs). As they are inherently resilient against variants, many works in the past have used these PRGs to perform effective detection. In essence, such works cast malware detection as a graph classification problem and apply existing graph mining and classification techniques \cite{Adagio,AppContext,DroidSift,droidol,fcgmal,hadm}. 

The efficacy of such approaches depends primarily on the features that they extract from PRGs. Prominent robust approaches from literature have used a variety of features such as API sequences \cite{AppContext,DroidMiner,DroidSift}, permissions used \cite{Drebin,droidapiminer}, information flows observed \cite{Mudflow}, instruction sequences used \cite{Adagio} and Control Flow Graph (CFG) patterns \cite{CSBD,HM} to learn discriminatory functions that could differentiate malware and benign apps. %As we explain in detail later in \S \ref{ss:mot2}, 
Understandably, each of these features represent a unique \textit{perspective} (interchangeably referred as \textit{view}) of apps, having their own merits and limitations. 

\subsection{Challenges in ML based malware detection}
In general, these ML and PRG based approaches are typically plagued by four challenges that make them unsuitable for large-scale malware detection in-the-wild:

\textbf{(C1) Expressiveness.} PRGs are known to be rich, complex and expressive data structures that characterize topological relationships among program entities. Representing them as vectors or other formats amenable for applying ML algorithms is a non-trivial task \cite{s&p}. In many cases, such representations fail to capture all the vital information from PRGs, thus losing their expressiveness, resulting in suboptimal detection rates \cite{cwlk,Adagio,hadm,fcgmal}. 
%For instance, solutions like \cite{mama}, \cite{DroidMiner}, \cite{Adagio} and \cite{massvet} capture the topological neighbourhood (i.e., structural) information from PRGs and detect security-sensitive behaviours through analysing them. However, as recent approaches \cite{AppContext} and \cite{DroidSift} revealed that an important contextual factor that distinguishes malice is whether or not the user is aware of such behaviours. Unfortunately, the above-mentioned methods which capture structural information well, fail to capture the contextual information and this leads to raising false alarms even when sensitive operations are performed with users’ consent. The general purpose graph kernels such as \cite{rw,frw,sp,graphlet,WLK,NHGK,NSPDK} also suffer with the same drawback.

\textbf{(C2) Scalability.} The scale of malware detection problem is such that we have millions of samples already and thousands streaming in every day. Many classic graph mining based approaches (e.g., \cite{s&p}) are NP hard and have severe scalability issues, making them impractical for malware detection in the wild \cite{massvet,parallelgk}.

\textbf{(C3) Integrating multiple views.} Each of the aforementioned feature sets (API sequences, information flows, etc.) provide a complementary view of the app, but no one view is completely sufficient to determine whether or not it is malicious. More specifically, some views are more amenable to detect certain attacks while they may not characterize other attacks well. For instance, information flow features are amenable for detecting 'privacy leak' attacks but are incapable of revealing 'privilege escalations'. A majority of existing approaches use either one or a selected few of the aforementioned views but not all of them. This prevents them from detecting a substantial majority of attacks. %In order to detect a substantial majority of attacks, an unified framework incorporating all the aforementioned features is required. Besides, the approach must be extendable in the sense that other views (e.g., dynamic analysis based views) could be easily added in the future without complicating the final result.

\textbf{(C4) Malicious code localization.}
 In general, almost all the ML based approaches  act as holistic black-box solutions as they just predict whether or not a given app (as a whole) is malicious (examples: \cite{Drebin,Adagio,AppContext,CSBD,Crowd,DroidMiner,DroidSift,HM,MLMalDetect,Mudflow,chabada,droidscribe,hadm,mama,prescience,reveal}). Such approaches are incapable of locating malice code portions (e.g., classes/methods containing such code). In general, precise malicious code localization would not only help to assess the approaches' trustworthiness but also cater several important applications such as supporting human analysts studying malware behaviors, engineering malware signatures and other counter-measures. Malicious code localization becomes particularly useful in the case of repackaged malware, as a predominant portion of their code remains benign and only a small portion is attack related. 

\subsection{Our Approach}
We take these four challenges into consideration and propose \tool{}, a unified framework which integrates all the above mentioned five views, namely, API dependencies, permission dependencies, information source and sink dependencies, instruction sequences and CFG patterns  to perform effective detection. Furthermore, leveraging on its multi-view analysis, it precisely locates malicious code portions in apps. In particular, \tool{} is developed with the three following design goals:

\textbf{1. Accuracy.} Accuracy of \tool{}, which is a multi-view PRG based approach depends on two factors: (i) how well it retains PRG’s expressiveness and (ii) how effectively it integrates different views.  In other words, it depends on how well challenges C1 and C3 are addressed. To address C1, we leverage on our previous work \cite{cwlk} and use the Contextual Weisfeiler-Lehman Kernel (CWLK) that is specifically designed to perform accurate malware detection by capturing both structural and contextual information from PRGs. 
% CWLK is demonstrated to retain PRGs’ expressiveness maximally and achieve excellent detection accuracies. 
To address C3, we resort to Multiple Kernel Learning (MKL) \cite{mklsurvey}, a well-known principled approach to integrate multiple feature sets with different modalities. To the best of our knowledge, ours is the first approach that captures contextual and structural information from different dependencies that emanate from PRGs representing different semantic views of an app.

\textbf{2. Efficiency.} \tool{} achieves its efficiency through the combined use of a scalable graph kernel (i.e., CWLK) and an efficient MKL approach, namely, Sequential Minimal Optimization (SMO) MKL \cite{SMO-MKL}. This addresses challenge C2.

\textbf{3. Locating malice code.} Besides effective detection, a salient feature of \tool{} is its ability to  locate malicious code portions in PRGs. We achieve this by meticulously extracting explainable features from each view and combining them in an interpretable manner. To this end, we propose a method for training a MKL Support Vector Machine (SVM) in the dual formulation and subsequently switching to primal formulation for prediction and interpretation. 
This process allows \tool{} to assign a maliciousness scores (m-scores, for short) to each node in the PRG which are then aggregated to arrive at m-scores of  methods and classes that encompass them. Again, to the best of our knowledge, ours is the first approach that performs such fine-grained malicious code localization without requiring any \textit{apriori} information on apps' piggybacking or composition. %Locating malice code snippets would be immensely helpful for human analysts as it reduces their search/inspection space from several thousands of methods/classes to a few tens.
This, in effect, addresses challenge C4.

\textbf{Experiments.} \tool{} is evaluated through large-scale experiments on more than 60,000 apps from benchmark datasets and collected in-the-wild. On benchmark datasets, \tool{} achieves more than 97\% F-measure, which is comparable to \soa{} approaches. More importantly, in an experiment closer to the real-world setting where the test-set is historically posterior (nearly by 2 years) to the training set, \tool{} achieves 71\% F-measure outperforming \soa{} techniques by 11\%. On recent wild apps, it achieves 72\% F-measure, outperforming  \soa{} approaches by 8\%. 
%In all the aforementioned experiments we observed that each of the individual feature sets produces sub-optimal F-measures and only through combining all of them using MKL, \tool{} achieves its superior F-measures. 
In all these experimental settings, \tool{} maintained better efficiency than two \soa{} techniques.
{\color{black} In malicious code localization experiments, our approach, on average, located more than 94\% of all malice classes within its top 10 classes with highest m-scores.} %None of existing approaches posses malicious code localization capabilities and hence are not comparable to \tool.

\textbf{Contributions.} In this work, we make the following contributions:

\begin{itemize} [leftmargin=*]
	\setlength\itemsep{0em}
	\item We leverage on CWLK, a graph kernel we proposed in our previous work \cite{cwlk} that is specially designed to perform malware detection by capturing both structural and contextual information from PRGs. Using CWLK we extract five different embeddings of an app's PRGs each representing a unique semantic view of the app.
	
	\item We propose \tool, a novel malware detection framework which systematically integrates these views using MKL thereby extracting a semantically richer representation. This helps \tool{} achieve superior accuracies over approaches which use either one or a selected few of the perspectives.

	\item We propose a kernel method based novel approach to locate and explain malicious code portions (i.e., methods/classes) in PRGs. To the best of our knowledge, ours is the first approach that performs automatic multi-view malicious code localization.
	
	\item We contribute to future research on malicious code localization by releasing \tool's results. M-scores for all basic blocks, methods and classes from apps in the benchmark datasets, \textsc{Drebin} \cite{Drebin} and \textsc{Mystique} \cite{mist} are made available at: \cite{oursite}.
	
\end{itemize}

%\textbf{Organization.}
The remainder paper is organized as follows: We begin by introducing the background and motivations for our framework’s design in \S \ref{sec:bgm} . The proposed \tool{} framework is presented in \S \ref{sec:meth}. The experimental design and implementation details are furnished in \S \ref{sec:edi}. Evaluation results and discussions are presented in section \S \ref{sec:rd}. Related work, limitations and conclusions are provided in \S \ref{sec:rw}, \S \ref{sec:lim} and \S \ref{sec:conc}, respectively.

\section {Background and Motivations}
\label{sec:bgm}

In this section, the background on kernel methods, Android malware detection and motivations for the two main components of the \tool{} framework, namely, CWLK and MKL are presented. %In particular, we begin by providing the background on kernel methods and then discuss the two following motivations in the two following subsections: (1)  and (2) why using only one semantic perspective is insufficient to detect different types of attacks and how integrating multiple such perspectives which complement each other would result in better detection accuracies. 

\subsection{Kernel Methods and Graph Kernels}
\label{ss:kmgk}
Kernel methods have been highly successful in solving a specific class of problems where feature vector representations of samples are not readily obtainable. Malware detection using PRGs is one of such problems. For many well-known classifiers, the data samples have to be explicitly represented as feature vectors through a user-specified feature map, $\phi$. In contrast, kernel methods require only a user-specified kernel $k$, i.e., a similarity function over pairs of samples in their native representations. Kernel methods work by mapping the samples into a feature space, implicitly and finding an appropriate decision boundary in the new feature space. Here, feature map $\phi(\cdot)$ is realized through the kernel function $k$, which facilitates computing inner products in the feature space using the samples in their native representation, i.e., for a pair of samples $x_i$ and $x_j$, $k(x_i, x_j ) = \langle \phi (x_i), \phi (x_j) \rangle$. 
%The kernel function must be positive definite for most kernel classifiers. Examples of positive definite kernels are the Dirac, Histogram Intersection and Gaussian kernels.
%Roughly speaking, \textit{a kernel value is a measure of similarity between a pair of samples}. %Therefore, a graph kernel is a similarity measure between graphs.

\textbf{Graph Kernels.}
Formally, a graph kernel $k : \mathbb{G} \times \mathbb{G} \rightarrow R$ is a kernel function defined on a domain of graphs, $\mathbb{G}$. Hence, given a labeled graph dataset $D_g = {(g_1, y_1),..., (g_n, y_n)}$ and a graph kernel $k$, a kernel classifier (e.g., Support Vector Machine (SVM)) can be directly used to perform graph classification. Graph kernels usually belong to the family of \textit{R convolution kernels}. These kernels decompose graphs into substructures such as walks, subgraphs etc. The comparison of two graphs is then based on the similarity between all pairs of such substructures. Several graph kernels have been proposed based on this idea \cite{rw,sp,WLK,NHGK}. %, for instance,\textit{ random walks kernels, shortest paths kernels} and Weisfeiler-Lehman (WL) kernels \cite{WLK}.

%Usually, the choice of a graph kernel for a particular task depends on three factors:\\
%\textbf{Scalability.} Since graphs are complex data structures with node/edge labels and attributes computing similarity among their sub-structures is often not scalable. For instance, classical graph kernels such as random-walk kernels, despite their nice theoretical properties, are computationally expensive making them unsuitable for a problem like malware detection where we process large volumes of graphs with thousands of nodes and edges. On the other hand scalable graph kernels such as WL and LTGK have been successfully used for malware detection at scale in the past.

\textbf{Explicit vs. Implicit Feature mapping.} Existing graph kernels can be classified into approaches that use explicit feature mapping ($\phi$) and those that directly compute a kernel function (i.e., $\phi$ is not necessarily known and may be of infinite dimension) \cite{gkunify}. Examples of former category include Weisfeiler-Lehman Kernel (WLK) \cite{WLK}, Neighborhood Hashing Graph Kernel (NHGK) \cite{NHGK} and CWLK, and that of latter category are Random Walk (RW) \cite{rw} and Shortest Path (SP) \cite{sp} kernels. Kernels that support explicit feature mapping have two advantages that make them particularly suitable for the malware detection problem:\\ 
\textit{(1) Scalability.} If explicit representations are manageable, these approaches usually outperform other kernels regarding runtime on large datasets, since the number of vector representations scales linear with the dataset size \cite{WLK,gkunify}.\\ 
\textit{(2) Explainability.} These kernels support extracting substructures of graphs as features and building a vocabulary of such features. This facilitates building explicit feature vector representation of individual graphs \cite{dgk,WLK}.
% Therefore, both kernel and non-kernel based classifiers could be used for classification. 
This aspect makes this category of kernels amenable for performing explainable malware detection\cite{gkunify}.

%Two explicit feature mapping kernels, namely, WLK \cite{WLK} and NHGK \cite{NHGK} have been successfully used for Android malware detection in \cite{MLMalDetect} and \cite{Adagio} respectively. Moreover, \cite{Adagio} performed explainable detection leveraging on the explicit feature map produced by NHGK.

%\textbf{Incorporating context.} However, these kernels only support incorporating node labels into features and not additional key node attributes such as context information. This motivates us to develop a new graph kernel that provides explicit feature mapping, scalability and also supports incorporating contextual information. 

CWLK, being a kernel with explicit feature maps achieves both high scalability and building explainable representations.

\subsection{Motivations for CWLK}
\label{ss:mot1}

\begin{figure*}[t]
	\centering
	\includegraphics[height=3cm,width=15cm]{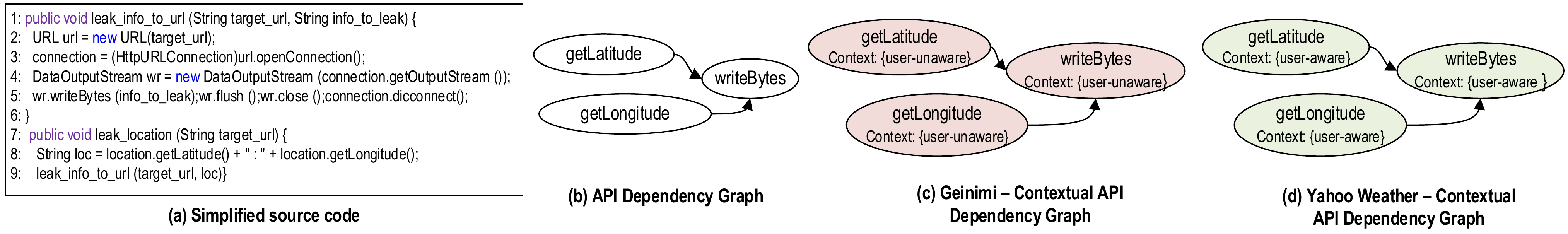}
	\caption{\small Location information being leaked in \textit{Geinimi} (malware) and \textit{Yahoo Weather} (benign) apps. (a) code snippet corresponding to leaking location information in both the apps. (b)  API Dependency Graph (ADG) corresponding to the location leak. (c) \textit{Geinimi's} Contextual ADG illustrating that it leaks information without the user's knowledge. (d) \textit{Yahoo Weather's} Contextual ADG illustrating that it leaks information with the user's knowledge. \label {fig:why-cwlk}}
\end{figure*}

This subsection explains with an example why considering structural information alone from PRGs is insufficient to determine the maliciousness of a sample and how supplementing it with contextual information helps to increase  detection accuracy.
To motivate this necessity, we use a primitive variant of malware from the \textit{Geinimi} family which steals users’ private information and contrast its behavior with that of a well-known benign app, \textit{Yahoo Weather}.

\textbf{\textit{Geinimi’s} execution.} The app is launched through a background event such as receiving a SMS or call. Once launched, it reads the user’s personal information such as geographic location and leaks the same to a remote server. The (simplified) malicious code portion pertaining to the location information leak is shown in Fig. \ref{fig:why-cwlk} (a). The method {\tt leak\_location} reads the geographic location through {\tt getLatitude} and {\tt getLongitude} APIs. Subsequently, it calls {\tt leak\_info\_to\_url} method to leak the location details (through {\tt DataOutputStream.writeBytes}) to a specific server. The API dependency graph (ADG)\footnote{The detailed procedure for constructing the ADG is provided later in \S \ref{ss:sa}.} corresponding to the code snippet is shown in Fig. \ref{fig:why-cwlk} (b). The nodes in ADG are labeled with the sensitive APIs that they invoke and the edges denote the control-flow dependencies. 

\textbf{\textit{Yahoo Weather’s} execution.} On the other hand, \textit{Yahoo Weather} could be launched only by user’s interaction with the device (e.g., by clicking the app’s icon on the dash board). The app then reads the user’s location and sends the same to its weather server to retrieve location-specific weather predictions. Hence, ADG portions of \textit{Yahoo Weather} is same as that of \textit{Geinimi}.

\textbf{Contextual information.} From the explanations above, it is clear that both the apps leak the same information in the same fashion. However, what makes \textit{Geinimi} malicious is the fact that its leak happens without the user’s consent. In other words, unlike \textit{Yahoo Weather}, \textit{Geinimi} leaks private information through an event which is not triggered by user’s interaction. We refer to this as a leak happening in \textit{user-unaware} context. On the same lines, we refer to\textit{ Yahoo Weather’s} leak as happening in \textit{user-aware} context. As explained in \cite{AppContext} and \cite{DroidSift}, one could determine whether a PRG node is reachable under \textit{user-aware} or \textit{user-unaware} context by examining its entry point nodes. Following this procedure we identify and add the context as an attribute to every ADG node to obtain the contextual ADG (CAGD). CADGs of \textit{Geinimi} \textit{Yahoo Weather} are shown in Fig. \ref{fig:why-cwlk} (c) and (d), respectively.

\textbf{Requirements for effective detection.} From the aforementioned example the two key requirements that make a malware detection process effective can be identified: \\
\textbf{(R1) Capturing structural information.} Since malicious behaviors often span across multiple nodes in PRGs, just considering individual nodes (and their attributes) in isolation is not enough. For instance, in the case of \textit{Geinimi}, the privacy leak attack spans across three ADG nodes. Therefore, capturing the structural (i.e., neighborhood) information from PRGs is of paramount importance. \\
\textbf{(R2) Capturing contextual information.} Considering just the structural information without the context is not enough to determine whether a sensitive behavior is triggered with or without users’ knowledge. For instance, if structural information alone is considered, the features of both \textit{Geinimi} and \textit{Yahoo Weather} apps become identical, thus making the latter a false positive. Hence, it is important for the detection process to capture the contextual information as well to make it more precise.

Many existing graph kernels could address the first requirement well. However, the second requirement which is more domain-specific makes the problem particularly challenging. To the best of our knowledge, the only graph kernel that addresses our two-fold requirement is CWLK \cite{cwlk}. Hence, we intend to use CWLK in this work for capturing both the aforementioned types of information from PRGs.

\subsection{Motivations for MKL}
\label{ss:mot2}

\begin{figure*}
	\centering
	\includegraphics[height=15cm,width=15cm]{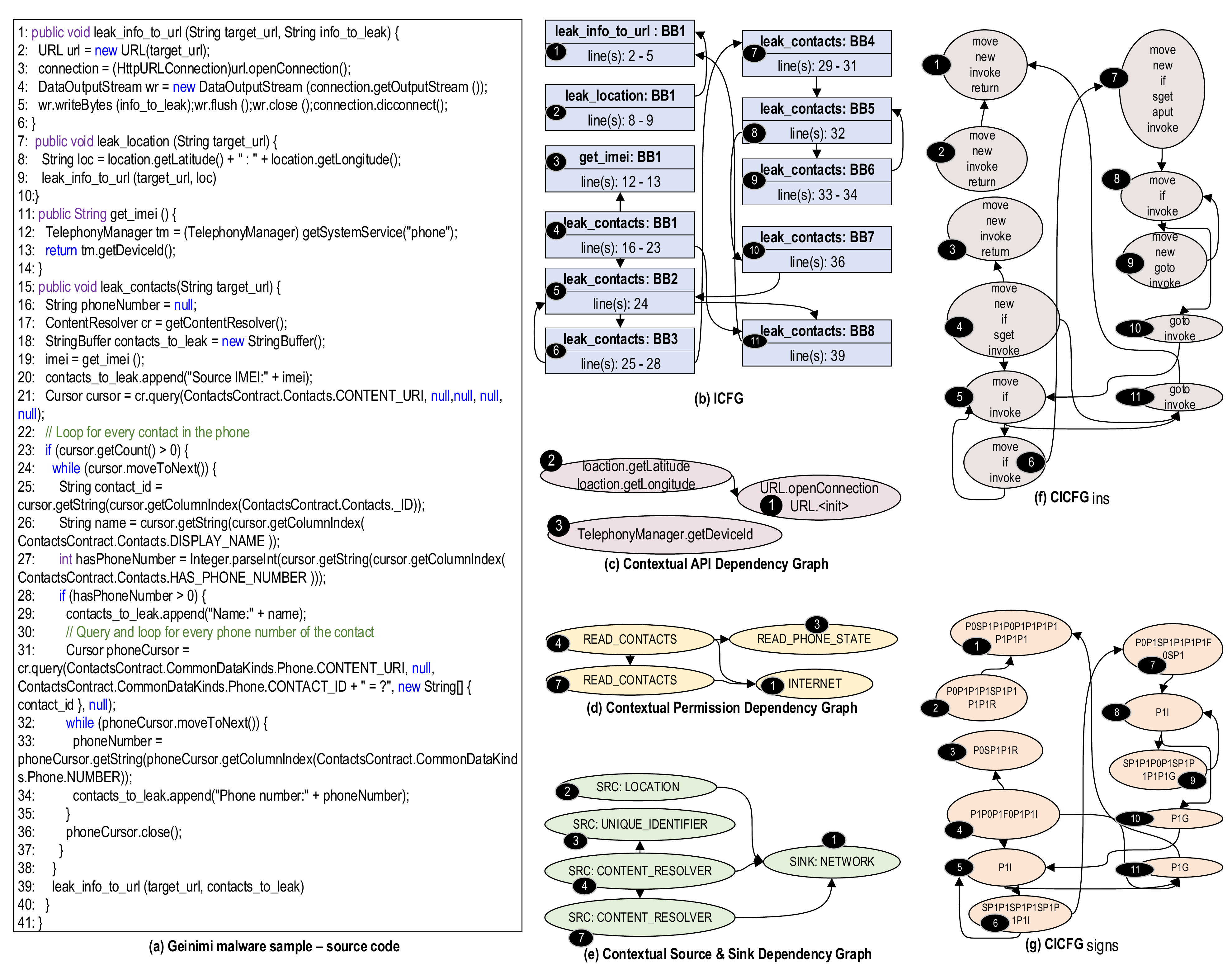}
	\caption{Extracting different five semantic views of the \textit{Geinimi} variant. (a) code snippet corresponding to leaking location, IMEI and contacts. (b) ICFG representation of the code. (c) Contextual API Dependency Graph corresponding to the code snippet. (d) Contextual Permission Dependency Graph. (e) Contextual Source \& Sink Dependency Graph. (f) Contextual ICFG with Dalvik bytecode instructions used as node labels. (g) Contextual ICFG with textual CFG signatures used as node labels. \label {fig:multiview}}
\end{figure*}
To illustrate the necessity to integrate multiple views, we consider a complex and more recent variant from the \textit{Geinimi} family. Besides leaking the geographic location, this variant leaks other  sensitive information such as device identifiers and contacts. 
%Since the importance of capturing the contextual information has already been illustrated, we refrain from discussing contexts in this subsection. 
The (simplified) malicious code portion that is responsible for these activities is shown in Fig. \ref{fig:multiview} (a). Firstly, the malware reads the geographic location and sends this information to a remote server, which acts as the command-and-control (C\&C) server, thereafter (lines 8-9). Subsequently, the malware reads the users’ contacts (i.e., names and phone numbers) from the content provider database and sends it to the C\&C server along with the IMEI number that serves as a unique identifier for the victimized device (lines 16-39). This malicious privacy leak comprises of a variety of actions that involves using APIs and URIs that are related to permissions, information sources and sinks (viz. {\tt openConnection, getDeviceId, Contacts.CONTENT\_URI,  Phone.CONTENT\_URI}) and other security-sensitive APIs (viz. {\tt getLatitude, getLongitude, getContentResolver, query}).

Through statically analysing the app, the graphs in Fig. \ref{fig:multiview}  (b) to (g) are constructed. While the formal definitions and detailed explanation of these graphs are provided later in \S \ref{ss:sa}, brief explanations are provided here to motivate capturing multiple views of an app through them. 

The Inter-procedural CFG (ICFG) corresponding to the aforementioned code is shown in Fig. \ref{fig:multiview}  (b). The basic blocks (i.e., is a sequence of instructions in a method with only one entry point and one exit point which represents the largest piece of the program that is always executed altogether) of methods form the nodes of the ICFG and the edges denote the control flow across these nodes. The node headers indicate the method name and the basic block number. The node content specifies the line numbers of code instructions belonging to the basic block. Subsequently, the context under which each ICFG node is reached is determined and added as a node attribute to obtain the Contextual ICFG (CICFG). In this particular case, all ICFG nodes are reached in the \textit{user-unaware} context. In sum, CICFG represents the control flow structure of the code in a precise and granular manner along with the contextual information.
 
\textbf{Abstraction to arrive at multiple views.} Having, constructed CICFG of an app, one way to detect malicious behavior is to abstract the functionalities performed by  its nodes and subsequently, detect the contextual subgraph patterns that correspond to such behaviors. We precisely follow this approach and propose five different abstractions of CICFG, each of which represent a semantic view of the app. They are, CADG (mentioned previously in \S \ref{ss:mot1}), Contextual Permission Dependency Graph (CPDG), Contextual Source \& Sink Dependency Graph (CSSDG), CICFG\textsubscript{ins}: CICFG nodes labeled with the \textit{Dalvik} instructions that they access and CICFG\textsubscript{signs}: CICFG nodes labeled with structured control flow signatures. These graphs for the \textit{Geinimi} example are depicted in Fig. \ref{fig:multiview} (c) to (g).

The CADG in Fig. \ref{fig:multiview} (c), is obtained by considering only the CICFG nodes that access security-sensitive Android APIs. All other nodes are removed and paths that exists between these nodes in the CICFG conditionally become edges in CADG. The CADG nodes are labeled with the sensitive APIs that they access. Similarly, we take into consideration the permissions, information sources/sinks and Dalvik instructions accessed in every CICFG node to produce CPDG, CSSDG and CICFG\textsubscript{ins} respectively.
%\footnote{\scriptsize PScout \cite{PScout} provides a means to map Android APIs and URIs to permissions and SUSI \cite{Susi} provides a means to map Android APIs to information sources and sinks. Using these two mappings, we deduce the permissions and the information sources/sinks used by an ICFG node.}. 
%For producing CICFG\textsubscript{ins}, the CICFG nodes are labeled with the Dalvik bytcode instructions that they access. All the remaining information such as instruction operands, APIs and parameters are removed.  
Lastly, CICFG\textsubscript{signs} is obtained by labeling every CICFG node with a textual signature that represents basic block's control flow structure. A control-flow analysis encoding grammar proposed by Cesare and Xiang in \cite{CFGGrammar} is used to infer these textual signatures.

\textbf{Complementary nature of the views.} With all these graphs constructed, it could be seen from Fig.\ref{fig:multiview} (c) that considering only the API dependencies, we could capture the behavior corresponding to leaking location information and IMEI number. However, behavior corresponding to leaking the contacts could not be detected. Similarly, considering only the permission dependencies (see Fig.\ref{fig:multiview} (d)), we could capture the behaviour corresponding to leaking the contacts but not the one corresponding to location information. Considering the dependencies among information sources and sinks, captures both the aforementioned leaks but fails to capture leaking IMEI (see Fig.\ref{fig:multiview} (e)). Also, CSSDG may not capture common malicious behaviours such as gaining root access or installing additional apps as they do not involve sources or sinks \cite{Mudflow}.
%All the three aforementioned graphs highly abstract the ICFG. ICFG-ins and ICFG-signs mitigates this by capturing the information at a much lesser granularity.
CICFG\textsubscript{ins} characterizes apps through the structural dependency among \textit{Dalvik} instructions. For instance, the privacy leaks in the example is characterized in a sequence involving nodes with {\tt move} and {\tt invoke} instructions. CICFG\textsubscript{signs} characterizes apps based on the control-flow structure of the code. This helps to model unusual control flow jumps, heavy usage of junk/unwanted instructions and complex loops which are predominant in malware \cite{CSBD,CFGGrammar}. 

In summary, each of these graphs represent different perspectives of a given app, capturing information at different levels of granularity with different modalities. While, each of these perspectives are capable of capturing certain characteristics of malware, they fail to capture certain other important characteristics due to their inherent limitations. 
%Individual perspectives could only reveal certain portions of malicious behaviour. 
We hypothesize that a much comprehensive and accurate malware detection model could be constructed by appropriately combining information from all these perspectives. 
Driven by this motivation, we construct a unified malware detection framework that is able to systematically integrate all the aforementioned views in the hope that, \textit{while a malware can disguise itself in some views, disguising itself in every view while maintaining malice intent will prove to be more strenuous}.

\section {Methodology}
\label{sec:meth}
The methodology of \tool{} framework designed to perform multi-views, context-aware malware detection and malicious code localization is presented in this section. We begin by describing the framework overview and subsequently, explain each component of the framework in separate subsections.

\begin{figure*}
	\centering
	\includegraphics[height=6cm,width=15cm]{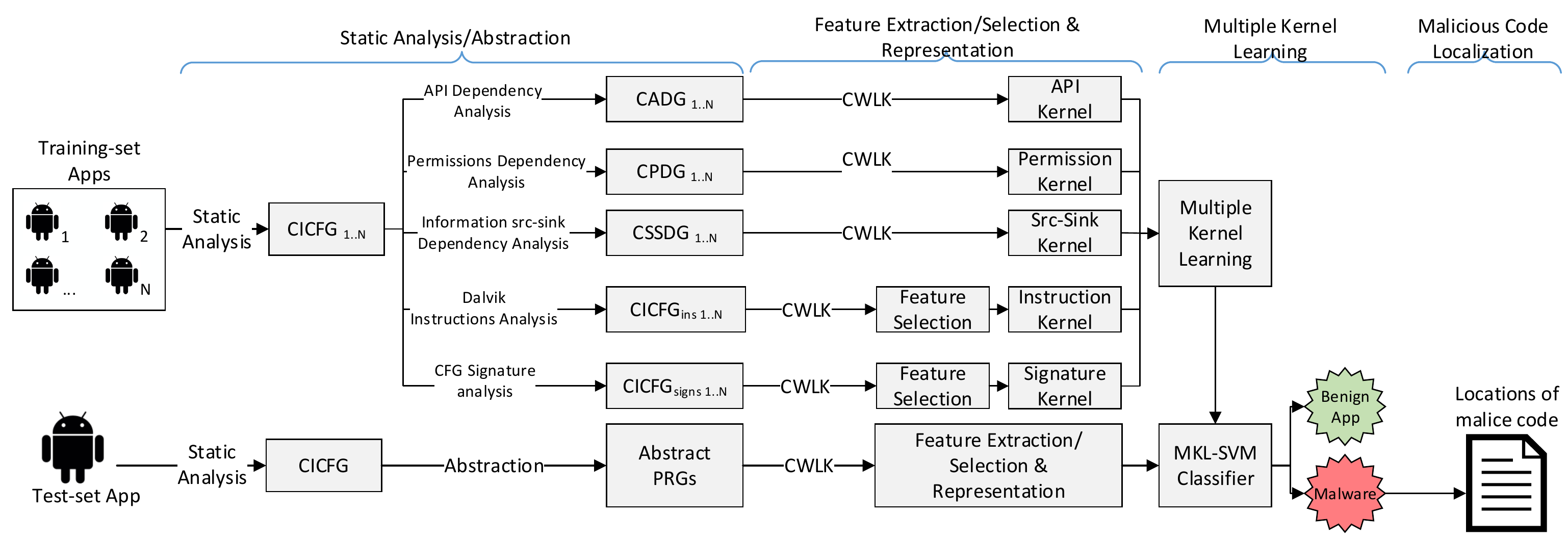}
	\caption{\tool{} - Overview \label{fig:ov}}
\end{figure*}

\subsection{\tool{} – Framework Overview}
\label{ss:ov}
As with any ML based framework, \tool{} has a separate model building (training) phase and evaluation (testing) phase. In the training phase, a set of known malware and benign apps are used to build the detection model. In the testing phase, the model is evaluated for its capability to automatically detect previously unseen malware. The overview of \tool{} framework is presented in Fig. \ref{fig:ov}. The framework has four components as described below.

\textbf{1. Static Analysis.}
To begin with, we perform static analysis on a given app and construct its CICFG. Subsequently, we construct the five above mentioned PRGs, namely, CADG, CPDG, CSSDG, CICFG\textsubscript{ins} and CICFG\textsubscript{signs} each of which represent a unique perspective of the app. This procedure is explained in detail in \S \ref{ss:sa}. 

\textbf{2. Feature Extraction, Selection and Representation.} 
Our framework considers contextual subgraphs from these five PRGs as semantic features to perform malware detection.
Hence, after these PRGs are constructed, those subgraphs which represent  security-sensitive events that happen in an app along with their context(s) are extracted as using CWLK \cite{cwlk}. CWLK yields separate feature vector representations for each PRG. In some cases when the dimensionality of these vectors are large, we use feature selection to reduce them. The detailed feature extraction, selection and representation procedure is explained in \S \ref{ss:fe}.

\textbf{3. Multiple Kernel Learning.}
After representing each of the PRGs as vectors, we need to appropriately combine them to build a single ML classifier that thoroughly leverages the strengths of individual views and remains robust to their weaknesses. In other words, the views have to be combined in a way that they complement each other and enhance the prediction accuracy. MKL \cite{mklsurvey}, provides a principled way to facilitate learning from this multi-view, multi-granular and multi-modal data. Therefore, we use SMO-MKL \cite{SMO-MKL}, a well-known MKL algorithm, to combine representations from all these PRGs and train an SVM to perform malware detection. The detailed procedure is explained in \S \ref{ss:mkl}. Subsequent to this training \tool{} is ready to perform malware detection at scale. 

\textbf{4. Malicious Code Localization.} During evaluation, if \tool{} predicts a sample to be malware, it further locates and reports the nodes in the CICFG (i.e., basic blocks) that perform malice operations and end up contributing significantly to the final prediction. That is, each CICFG node is assigned an m-score that quantifies the statistical significance of malice operation(s) it is involved in. The m-scores of basic blocks are aggregated to arrive at the same of their encompassing methods and classes. In most cases, nodes with high m-scores reveal the app's characteristics related to its malice behaviors. The detailed procedure is presented in \S \ref{ss:lm}.

Now, each components of \tool{} is explained in detail in the four following subsections.

\subsection{Static Analysis}
\label{ss:sa}

As a first step towards constructing the five above mentioned PRGs for a given app, we perform static control-flow analysis and construct its ICFG. Formally, an ICFG is defined as follows:

\textbf{Definition 1 (ICFG).} $ICFG = (N_i, E_i)$ for an app \textit{a} is a directed graph in which each node $bb \in N_i$ denotes a basic block of a method \textit{m} in \textit{a}, and each edge $e(bb_1, bb_2) \in E_i$ denotes either an intra-procedural control-flow dependence from $bb_1$ to $bb_2$ or a calling relationship from $bb_1$ to  $bb_2$ and $ E_i \subseteq N_i \times N_i$.

Compared to other well-known PRGs such as CG, ICFG is a  more fine-grained representation of the control flow sequence inside an app. Hence working on abstractions of ICFGs enable us to capture the finer details of apps, which are helpful in distinguishing malicious and benign behaviors with greater precision. Hence, we choose to abstract apps' behaviors from their ICFGs.

\textbf{CICFG construction.}
Once an app's ICFG is constructed, we proceed towards identifying contexts under which every ICFG node is reached to build its CICFG.
Several works such as DroidSIFT \cite{DroidSift}, AppContext \cite{AppContext} and Elish et al. \cite{karim} have proposed techniques to identify whether a PRG node is reached under the \textit{user-aware} or \textit{-unaware} context by analyzing their entry-points (i.e., nodes that do not have any incoming edge). We follow the approach mentioned in DroidSIFT \cite{DroidSift}.
Formally a CICFG, is defined as follows: 

\textbf{Definition 2 (CICFG).} $CICFG = (N_i, E_i, \xi)$ of an app $a$ is a directed graph in which each node $bb \in N_i$ denotes a basic block of a method $ m $ in $a$, and each edge $e(bb_1, bb_2) \in E_i$ denotes either an intra-procedural control-flow  or a calling relationship from $bb_1$ to $bb_2$. $\xi$ is a set of contexts through which every node $bb \in N_i$ could be reached.

%In our implementation, Androguard \cite{AG} is used for performing control-flow analysis on byecode of apps and constructing ICFGs.

Once the CICFG of an app is constructed, we abstract it with various Android platform specific analysis to construct the five different PRGs that \tool{} leverages on. For each of the PRGs, the procedure to construct them and their formal definitions are provided along with relevant explanations below.

\textbf{CADG construction.} Intuitively, CADG of an app is obtained from its CICFG by considering only the nodes that access security-sensitive APIs\footnote{Two existing works, PScout \cite{PScout} and SUSI \cite{Susi} list commonly known security-sensitive Android APIs.  We use these two lists to identify sensitive APIs.}. All other nodes are removed and paths that exists between such nodes in the CICFG become edges in CADG, if they satisfy certain conditions as described in the definition below. If multiple APIs are invoked in a single CICFG node, it is labeled with the sorted list of unique APIs being invoked. CADG's formal definition is as follows:

\textbf{Definition 3 (CADG).}
CADG can be represented as a 4-tuple, $CADG = (N_A, E_A, \lambda_A, \xi)$, where $N_A$ is a finite set of nodes and $n \in N_A$ is a basic block that accesses at least one security-sensitive API. $E_A \subseteq N_A \times N_A$  is a set of directed edges where an edge from $e (n_1, n_2) \in E$ exists, iff there exists a path $p (n_1, n_2)$ between these two nodes in the CICFG and $ method(n_1) $ = $method (n_2) $, where $ method(n) $ denotes the method that encompasses basic block $n$\footnote{This follows from the observation that in most malware the malice code portion is closely-knit i.e., spanning only to a few methods. We also attempted two other variants of CADG. We reduce the path in CICFG to edges in CADG (i) only if the calling and called nodes belong to the same package and (ii) only if the calling and called nodes belong to the same class. Both these variants contained much larger number of edges and also failed to capture the attacks as effectively as the CADG defined above (experimentally verified).}. $\lambda_A$ is the set of labels representing the security-sensitive APIs and $\ell_A: N_A \rightarrow \lambda_A$ is a labeling function which assigns a label to each node. 
$\xi$ is a set of contexts through which every node in the CADG could be reached and $\mathcal{C} \rightarrow \xi$ is a function which assigns the context to each node.

%\textit{Security-sensitive APIs}:  A security-sensitive API is a method that meets at least one of the following three requirements: (1) Permission-protected methods. Some methods and Android APIs require permissions to be invoked. Such methods usually access security-sensitive resources and data (the detailed list of the methods is specified in PScout \cite{PScout}). (2) API that is either a source or a sink of an information flow. An information flow consists of a source from which the security-sensitive data may originate and a sink to which the data may be sent (the detailed list of sources and sinks are specified in SUSI \cite{Susi}). Sources and sinks are not always protected by permission and similarly a permission-protected method may not always access a source/sink \cite{AppContext} (3) Reflection methods [], dynamic [] and native code-loading methods []. Resolving reflection, native and dynamic loading methods in static analysis is a known difficult problem with fundamental limitations \cite{AppContext}. For this reason, we do not attempt to resolve these methods, but rather treat them as being security sensitive. In doing so, we are being conservative, because these methods may result in invoking other security-sensitive methods. There are a few methods in the Android API allowing apps to load and invoke code at runtime that has also been leveraged by existing malware.

\textbf{CPDG.} Similarly, CPDG of an app is obtained from its CICFG by considering only the nodes whose functionality pertains to using Android permission(s). APIs and URIs observed in  nodes are used to determine their permissions pertinence\footnote{PScout \cite{PScout} provides a mapping from Android APIs and URIs to permissions required to access them. Furthermore, we infer the usage of \textit{intents}, \textit{reflection} and \textit{native code} through relevant APIs and consider them as using special permissions. We use these mappings to build CPDGs.}. All other nodes are removed and paths that exists between such nodes in the CICFG conditionally become edges in CPDG. If multiple permissions are used in a single CICFG node, then the same strategy followed in CADG is used to label corresponding CPDG nodes.

\textbf{Definition 4 (CPDG).}
Formally, $CPDG = (N_P, E_P, \lambda_P, \xi)$, and node $n \in N_P$ is a basic block whose functionality pertains to using permissions. $E_P \subseteq N_P \times N_P$  is a set of directed edges where an edge from $e (n_1, n_2) \in E_P$ exists, iff there exists a path $p (n_1, n_2)$ between these two nodes in the CICFG and $ method(n_1) $ = $method (n_2) $. $\lambda_P$ is the set of labels representing the concerned permission(s) and $\xi$ is a set of contexts through which every node in the CPDG could be reached.

\textbf{CSSDG.} 
CSSDG of an app is obtained from its CICFG by considering only the nodes whose functionality pertains\footnote{To identify information sources and sinks accessed in CICFG nodes, we leverage on SUSI \cite{Susi} and  MUDFLOW \cite{Mudflow}. Together, these works map Android APIs and URIs to 15 source and 18 sink categories.} to using information sources (e.g., contacts) and sinks (e.g., network). All other nodes are removed and paths that exists between such nodes in the CICFG conditionally become edges in CSSDG. %If multiple sources/sinks are used in a single CICFG node, the same strategy followed in CADG is used label corresponding CSSDG nodes.

\textbf{Definition 5 (CSSDG).}  
Formally, $CSSDG = (N_S, E_S, \lambda_S, \xi)$, and node $n \in N_S$ is a basic block whose functionality pertains to information sources or sinks. $E_S \subseteq N_S \times N_S$  is a set of directed edges where an edge from $e (n_1, n_2) \in E_S$ exists, iff there exists a path $p (n_1, n_2)$ between these two nodes in the CICFG and $ method(n_1) $ = $method (n_2) $. $\lambda_S$ is the set of labels representing the concerned source(s) or sink(s) and $\xi$ is a set of contexts through which every node could be reached.

\textbf{CICFG\textsubscript{ins}.} Recently, studies such as \textsc{Adagio} \cite{Adagio} demonstrated that structural  information from PRGs labeled with \textit{Dalvik} instruction categories (e.g., {\tt move, add, iget}, etc.) could capture security-sensitive behaviors and thus help detecting malware effectively. Inspired by them, we extend this approach by supplementing the instruction-level structural information with contextual information. To this end, we build CICFG\textsubscript{ins} as described below.

\textbf{Definition 6 (CICFG\textsubscript{ins})} Formally, $CICFG_{ins} = (N_i, E_i, \lambda_I, \xi)$, where $N_i$, $E_i$  and $\xi$ are the nodes, edges and node contexts in the CICFG, respectively. The function $\ell_i: N_i \rightarrow \lambda_I$ labels every node $n \in N_i$ with the categories of \textit{Dalvik} instructions\footnote{To determine the categories of Dalvik instructions to be used as CICFG\textsubscript{ins} node labels, we refer to \textsc{Adagio} \cite{Adagio}. The authors manually analyzed and categorized all the instructions into 15 distinct categories (such as {\tt move}, {\tt invoke}, etc.).}  that it accesses.

\textbf{CICFG\textsubscript{signs}.} 
{\color {black} Similar to \textsc{Adagio}, Allix et. al. \cite{CSBD} proposed an approach which leverages on control-flow structural information. They represent the structure of basic blocks in every method as textual signatures following a method devised by Pouik et al. \cite{AGClone}\footnote{Pouik et al. \cite{AGClone} leveraged on a grammar proposed by Cesare and Xiang  \cite{CFGGrammar} to represent CFG textual signatures in their work on establishing similarity between Android apps.}. 
This signature is an abstraction of code's structure, but discards low-level details such as variable/register names and numbers. This property is particularly desirable for malware detection as variants from same family may share the same abstract CFG while having different bytecode. Also, this helps to model unusual control flow structure such as jumps, heavy usage of junk/unwanted instructions and complex loops which are \textit{tell-tale} signs of malware \cite{HM,Adagio,CSBD,AG}. Overall, using an abstract signature representation of CFG basic blocks could allow taming common obfuscations used by malware. 
Inspired by Allix et. al's approach, we extend it by supplementing the CFG signature-level structural information with contextual information. To this end, we build CICFG\textsubscript{signs} as described below. }

\textbf{Definition 7 (CICFG\textsubscript{signs})} 
Formally, $CICFG_{signs} = (N_i, E_i, \lambda_s, \xi)$, where $N_i$, $E_i$ and $\xi$ are the nodes, edges and contexts in the CICFG, respectively. The function $\ell_s: N \rightarrow \lambda_s$ labels every node $n \in N_i$ with the control-flow signatures arrived at using  Cesare and Xiang’s grammar \cite{CFGGrammar}. 

\subsection{Feature Extraction, Selection and Representation}
\label{ss:fe}
Once the five PRGs are constructed as described above, we proceed to extract and select contextual subgraph features from each of them using CWLK and represent them as vectors.

\textbf{Feature Extraction using CWLK.} 
CWLK, a graph kernel developed in our previous work \cite{cwlk} is specifically designed to cater effective malware detection by capturing both structural and contextual information from PRGs. This directly addresses the requirements R1 and R2 stated in \ref{ss:mot1}. Since CWLK could be used with any type of PRG, we explain its working in general irrespective of the PRG type.
It is used in the same manner to represent all our PRGs as vectors.

%Upon labeling nodes in the CICFG, each basic block is characterized by the APIs, permissions, Dalvik instructions and CFG structure in CADG, CPDG, CICFG\textsubscript{ins} and CICFG\textsubscript{signs} respectively. However, our method strives to model the context of each basic block (i.e. a node in the aforementioned graphs) and thus the neighbourhood of a basic block must be taken into account. This is equivalent to considering the context in which a particular characteristic is exhibited by an app. For instance, the context in which a particular API is invoked or a permission is being used is modelled by considering the neighbourhoods of nodes on ADG and PeDG, respectively. To this end, for each node in each of the graphs, we compute a neighbourhood-based label (over all of its neighbours upto degree 3), using the WL graph kernel originally proposed by Shervashidze et. al [] and further compute kernel between pairs of graphs.

%CWLK belongs to the family of convolution kernels, which is based on the idea of the decomposition of graphs into subgraphs in such a way that a kernel function for a pair of graphs can be defined as a convolution of kernel functions defined over their subgraphs. 

The main idea behind CWLK is to condense the structural and contextual information contained in a PRG neighborhood into a single label value.
CWLK computes the similarities between a given pair of PRGs $ G = (N, E, \lambda, \xi) $ and $ G' = (N',E',\lambda, \xi) $ based on the 1-dimensional WL test of graph isomorphism \cite{WLK}. The algorithm works by iteratively augmenting the node labels by the sorted set of labels of neighboring nodes along with their context(s). 
This label-enrichment process is referred as \textit{contextual relabeling} and new labels are referred as \textit{ contextual neighborhood labels}. 
Thus, in each iteration \textit{i} of the algorithm, for each node $n \in N$, we get a new contextual neighborhood label, $\gamma_{i}(n)$ that encompass the $i^{th}$ degree neighborhood around $n$ and along with \textit{n}'s context. This characterizes that the neighborhood, $\gamma_{i}(n)$ could be reached under the context, $ \xi(n) $. 
For graph $G$, this contextual relabeling process yields a Contextual WL (CWL) graph at height $i$, denoted as $\mathcal{G}_i = (N,E,\gamma)$. Thus for any given graph $G$, we could obtain a sequence of CWL graphs as defined below.

\textbf{Definition 8 (CWL sequence).} Define the CWL graph at height $i$ of the graph $G = (N,E,\lambda, \xi)$ as the graph $\mathcal{G}_i = (N,E,\gamma_i)$.
The sequence of graphs 
\begin{equation}
{\mathcal{G}_0, \mathcal{G}_1, ..., \mathcal{G}_h} = {(N,E,\gamma_0),(N,E,\gamma_1), ...,(N,E,\gamma_h)}
\end{equation}
is called the CWL sequence up to height $h$ of $G$, where $\mathcal{G}_0 = G$ (i.e., $\gamma_0 = \lambda$) is the original graph and $\mathcal{G}_1 = r(\mathcal{G}_0)$ is the graph resulting from the first relabeling, and so on.

\textbf{Contextual Relabeling Algorithm.}
Since the contextual relabeling algorithm is the key step in computing CWLK value between a given pair of PRGs, we explain the same in detail through algorithm \ref{algo:cr} in Appendix \ref{app:cr}.

Once the CWL sequences for a pair of PRGs  are computed, the CWLK kernel over them is defined as follows:

\textbf{Definition 9 (CWLK).} Given a valid kernel $k(.,.)$ and the CWL sequence of graph of a pair of PRGs $G$ and $G'$, the contextual WL graph kernel with $h$ iterations is defined as 
\begin{equation}
\label{eq:cwlk}
k^{(h)}_{CWL}(G, G') = k(\mathcal{G}_0, \mathcal{G}'_0) + ... + k(\mathcal{G}_h, \mathcal{G}'_h)
\end{equation}
where $h$ is the number of CWL iterations and ${\mathcal{G}_0, \mathcal{G}_1, ..., \mathcal{G}_h}$ and ${\mathcal{G}'_0, \mathcal{G}'_1, ..., \mathcal{G}'_h}$ are the CWL sequences of $G$ and $G'$, respectively. 

\begin{figure}[t]
	\centering
	\includegraphics[height=3cm,width=9cm]{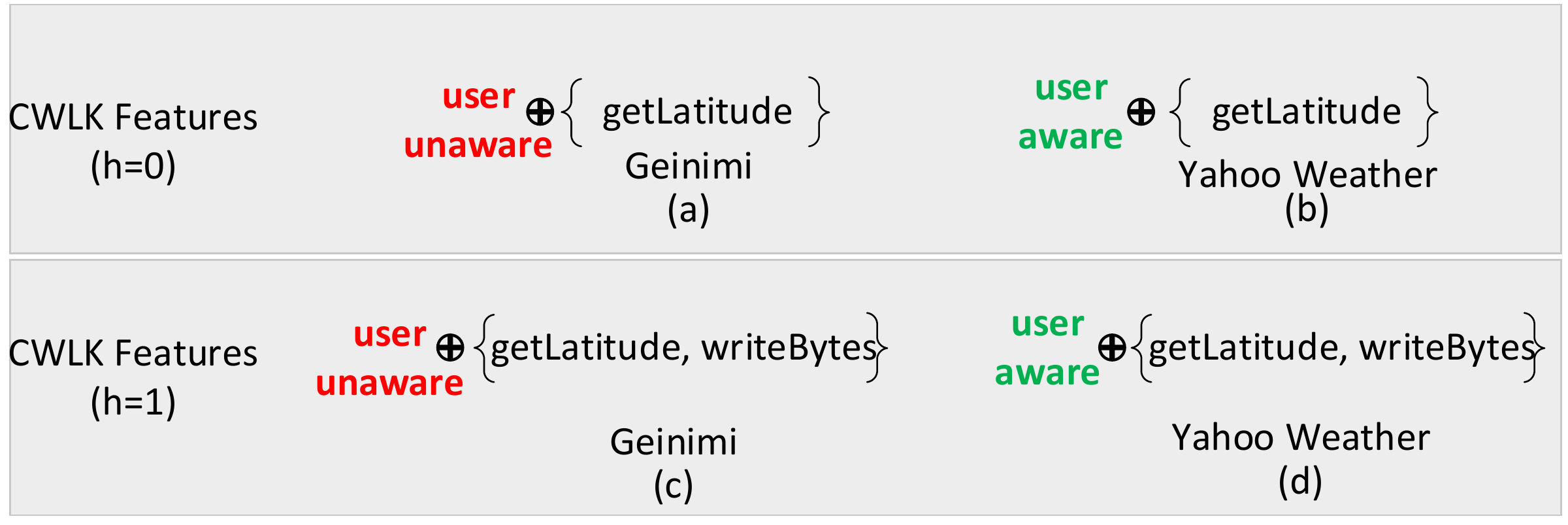}
	\caption{\small CWLK neighborhood labels for the node {\tt getLatitude} from \textit{Geinimi} and \textit{Yahoo Weather} apps. \label {fig:illus}}
\end{figure}
\textbf{Example of CWLK's working.}
Having presented the formulations for CWLK, we now  illustrate how it helps to detect malicious subgraphs regions from PRGs with an example. Lets consider the \textit{Geinimi} and \textit{Yahoo Weather} examples mentioned in \S \ref{ss:mot1}.
Applying CWLK on the CADGs of \textit{Geinimi} and \textit{Yahoo Weather} examples  (see Fig. \ref{fig:why-cwlk} (c) and (d)), for the node {\tt getLatitude}, for heights $h = 0, 1$, we get the neighborhood labels as shown in Fig. \ref{fig:illus} (a)-(d). 

In both cases, the node {\tt getLatitude} has only one degree-1 neighbor, {\tt writeBytes} and this fact is reflected in the neighborhood label.  Clearly, CWLK captures the structural information around the node {\tt getLatitude}, incrementally in every iteration of $h$.  In fact, neighborhood label for $h=1$ captures that a sensitive node, {\tt writeBytes} lies in the neighborhood of {\tt getLatitude}, which highlights a possible privacy leak (see Fig. \ref{fig:illus} (c) and (d)). However, looking at only the structural information, one cannot distinguish the \textit{Geinimi's} malicious and \textit{Yahoo Weather's} benign information leak. %It is noted this is the only information captured by WLK \cite{WLK} (and other general purpose graph kernels).

 As it is evident from the figure, besides capturing the composition of the neighborhood, CWLK also captures whether the neighborhood is reached in \textit{user-aware} or \textit{unaware} context. Clearly, the contextual neighborhood labels of \textit{Geinimi} reveal that the sensitive operations are performed in the \textit{user-unaware} context, unlike \textit{Yahoo Weather}.

Hence, it is evident that the CWLK’s contextual relabeling provides a means to appropriately distinguish malicious PRG neighborhoods from the benign ones, establishing its suitability for malware detection.

\textbf{CWLK time complexity and validity.}
The runtime complexity of CWLK with $h$ iterations on a graph with $n$ nodes and $e$ edges is $O(he)$ which is same as that of WLK. Meaning, capturing contextual information does not reduce the efficiency. For more information on derivation of CWLK's time complexity and proof of positive semi-definiteness, we refer the reader to the original work \cite{cwlk}.

\textbf{Explicit feature vector representation.} 
Give a dataset of $ K $ PRGs, CWLK uses Bag-of-Features (BoF) model to yield their feature vector representations (i.e., graph embeddings) and subsequently kernel matrix could be computed.
This process involves the following steps: 
\begin{itemize}
[leftmargin=*]
\setlength\itemsep{0em}
\item A vocabulary $\Sigma^*$ of all the contextual neighborhood labels of nodes across $K$ graphs is obtained. Thus each graph is represented as $|\Sigma^*|$ dimensional vector.
\item Subsequently, $K \times K$ kernel matrix can be computed from the dot product of these vectors. 
%In summary, CWLK has the same efficiency as WLK yet supports richer feature vector representations which retain CADGs'  expressiveness better.
\end{itemize}

\textbf{Feature selection.}
With the kernel matrix thus obtained, an MKL classifier could be trained to detect malware. However, we note that the vocabularies of contextual subgraph features, $\Sigma^*$, for CICFG\textsubscript{ins} and CICFG\textsubscript{signs}\footnote{The reason why such an issue rises only in the case of CICFG\textsubscript{ins} and CICFG\textsubscript{signs} is understandable. That is, in the case of CADG, CPDG and CSSDG, the number of unique node labels is limited by the  APIs, permissions, information source and sink categories available. Consequently, limited contextual neighborhood labels to emerge from the relabeling process and thereby limiting the size of the vocabulary. However, in the case of CICFG\textsubscript{ins} and CICFG\textsubscript{signs}, the number of unique node labels (i.e., the number of unique instruction sequence and CFG signatures, respectively) across the whole dataset is extremely large, leading to mammoth vocabulary $\Sigma^*$.} are extremely large (more than 500,000 features emerge from these views in all our experiments. See \S \ref{sec:rd} for more details). This leads to building very high dimensional embeddings which adversely affects both the accuracy and efficiency. 
Hence, to mitigate this, we perform feature selection over the CICFG\textsubscript{ins} and CICFG\textsubscript{signs} embeddings using the chi-squared feature selection algorithm \cite{chi} and then compute their respective kernel matrices. The number of features to be selected from these PRGs is empirically determined to be 5,000. This helps to retain only the informative subgraph features, thereby preventing overfitting and improving efficiency.

\subsection{Multiple Kernel Learning}
\label{ss:mkl}

Once the feature vectors of all the apps in the training-set are built for all five views, we train an MKL classifier. This procedure is explained below with relevant notations.

\textbf{Notations.} Denote the features of an app $\textbf{x}$ corresponding to each of views as following vectors:
CADG: $\overrightarrow{\textbf{x}}_a = [x^1_a, x^2_a,...]^{T}$, CPDG: $\overrightarrow{\textbf{x}}_{p} = [x^1_p, x^2_p,...]^{T}$, CSSDG: $\overrightarrow{\textbf{x}}_{ss} = [x^1_{ss}, x^{2}_{ss},...]^{T}$,
CICFG\textsubscript{ins}: $\overrightarrow{\textbf{x}}_{in} = [x^1_{in}, x^2_{in},...]^{T}$, {\tt Signature}: $\overrightarrow{\textbf{x}}_{si} = [x^1_{si}, x^2_{si},...]^{T}$
where $x^{i}_v$ denotes individual features emerging from view $v$ and the set of all the views is denoted as $V = \{a,p,ss,in,si\}$.
The label corresponding to an app $\textbf{x}^{(i)}$ is denoted as $y^{(i)} \in \{-1, +1\}$, where $-1$ indicates benign and $+1$ indicates malicious apps. Let the total number of apps in the training set be $K$.
The kernel value between a pair of apps $\textbf{x}^{(i)}$ and $\textbf{x}^{(j)}$ corresponding to each view is computed as follows:

\begin{equation}
k_{v} (\textbf{x}^{(i)},{\textbf{x}}^{(j)}) = \langle \overrightarrow{\textbf{x}}^{(i)}_v,\overrightarrow{\textbf{x}}^{(j)}_v \rangle 
\end{equation}
where $\overrightarrow{\textbf{x}}^{(i)}_v$ denotes the vector representation of   $\textbf{x}^{(i)}$ in  view $v$  and $\langle \cdot , \cdot \rangle$ denotes dot product over a pair of vectors.  Meaning, the similarity measured in view $v$ is nothing but the normalized linear kernel. 
Following this procedure, the kernel matrix across all the apps in the training set for each view is arrived.
For the sake of simplicity, we refer to the kernel built from CADG view as {\tt API} kernel. Similarly, the four remaining kernels are referred as {\tt Permission}, {\tt Src-sink}, {\tt Instruction} and {\tt Signature} kernels.
 %The kernel matrix for each of view in $V$ are denoted as $\textbf{k}_a, \textbf{k}_p,\textbf{k}_{ss}, \textbf{k}_{in}$ and $\textbf{k}_{si}$, respectively. These are also referred as \textit{base kernels}.

\textbf{Kernel methods.} Given a kernel matrix over the training samples, the goal of classical kernel-based learning with SVMs is to learn the vector, $\alpha$, describing each sample $ \textbf{x} $'s contribution to the hyperplane that separates the points of the two classes (\textit{aka} decision boundary) with a maximal margin \cite{SVMSurvey} and can be found with the following optimization problem:

\begin{equation}
\min_{\alpha} \bigg( \frac{1}{2} \sum_{i=1}^{K} \sum_{j=1}^{K}  \alpha^{(i)} \alpha^{(j)} y^{(i)} y^{(j)} k \big( \textbf{x}^{(i)},\textbf{x}^{(j)} \big) - \sum_{i=1}^{K} \alpha^{(i)} \bigg)
\end{equation}

subject to constraints,
\begin{equation}
\begin{split}
 \sum_{i=1}^{K} \alpha^{(i)} y^{(i)} = 0  \\
  0 \le \alpha^{(i)} \le C
\end{split}
\end{equation}
Eq. (5) constrains the $\alpha$’s to be non-negative and less than some constant $C$. $C$ allows for soft-margins, meaning that some of the examples may fall between the margins. This helps to prevent over-fitting the training data and allows for better generalization.

Given $\alpha$ found in eq. (4), we have the following decision function:

\begin{equation}
f(\textbf{x}) =  sign \bigg(\sum_{i=1}^K \alpha^{(i)} y^{(i)} k(\textbf{x}^{(i)},\textbf{x})\bigg)
\end{equation}
where the function $ sign $ returns $+1$ if the summation term is positive, and $-1$ otherwise. 

If there exist vectorial representations  $\overrightarrow{\textbf{x}}^{(i)}$ for each sample ${\textbf{x}}^{(i)}$ in the training set, then a vector $\overrightarrow{\textbf{w}}$ (called \textit{weight vector}) could be deduced such that,
\begin{equation}
\overrightarrow{\textbf{w}} =  \sum_{i=1}^K \alpha^{(i)}y^{(i)}\overrightarrow{\textbf{x}}^{(i)}	
\end{equation}
and eq. (6) could be written equivalently as,
\begin{equation}
f(\textbf{x}) =  sign \big( \langle \overrightarrow{\textbf{w}},\overrightarrow{\textbf{x}}\rangle \big) =  sign \big(\sum_{\mathfrak{f}=1}^{|\textbf{x}|} w^\mathfrak{f}x^\mathfrak{f} \big)
\end{equation}

It is noted that, in eq. (8), individual component of the weight vector $w^\mathfrak{f}$ denotes the weight (or relative importance) of feature $\mathfrak{f}$ and $x^\mathfrak{f}$ denotes the frequency of occurrence of $\mathfrak{f}$ in $\textbf{x}$.   
Alternatively, $f(\textbf{x})$ could be formulated as an optimization problem over $\textbf{w}$ and solved as follows:
\begin{equation}
\min_{\overrightarrow{\textbf{w}}} ||\overrightarrow{\textbf{w}}||^2 + \sum_{i=1}^K max(0,1-y^{(i)}f(\overrightarrow{\textbf{x}}^{(i)}))
\end{equation}

%The SMO-MKL classifier receives a number of samples, $x_i$, and their labels, $y_i$, and trains using this labeled data. Given a new unseen sample, $x$, the goal of CW classifier is to predict the label, $y$, of this new sample based on its trained model. 

\textbf{MKL.} With MKL, we are interested in finding $\beta$, in addition to the standard $\alpha$ of SVMs, such that

\begin{equation}
k_{comb} (\textbf{x}^{(i)},\textbf{x}^{(j)}) = \sum_{v \in V} \beta_v k_v(\textbf{x}^{(i)},\textbf{x}^{(j)})
\end{equation}
is a linear combination of all the kernels $v \in V$ with $\beta_v \ge 0$, where each kernel, $k_v$, uses a distinct set of features emanating from different views of apps \cite{mklsurvey}. The general outline of the algorithm is to first combine the kernels with $\beta_v = 1/|V|$, find $\alpha$, and then iteratively keep optimizing for $\beta$ and $\alpha$ until convergence. $\beta_v$ is also referred as the weight of kernel $v$, which quantifies the relative importance of view $v \in V$.

\textbf{SMO-MKL.}
To solve for kernel weights ($\beta$), and support vectors ($\alpha$), simultaneously, we use the SMO based MKL algorithm proposed in \cite{SMO-MKL}. For details on SMO optimization and subsequent computations of $\alpha$ and $\beta$, we refer the reader to the original work by Vishwanathan et. al. \cite{SMO-MKL}.

This method of MKL using the SMO algorithm is reported to be very efficient. Solving for $\beta$ and $\alpha$ with as many as 50,000 samples and 300,000 kernels has been shown to take just over 30 minutes on many applications from different domains such as Computer Vision and Bioinformatics \cite{SMO-MKL}. 
%It is important to note that this optimization problem only needs to be solved once, as the support vectors ($\alpha$) and kernel weights ($\beta$) found can be used to classify all the test-set samples. 
In the MKL context, eq. (6) used to predict the label of a given sample is realized as follows:

\begin{equation}
f(\textbf{x}) = sign \bigg( \sum_{i=1}^K \alpha^{(i)} y^{(i)} k_{comb}(\textbf{x}^{(i)},\textbf{x}) \bigg)
\end{equation}

Finding the kernel weights and support vectors  across all views culminates the training process, yielding an MKL-SVM ready to perform multi-view malware detection. 

\subsection{Malicious code localization}
\label{ss:lm}
Once the MKL-SVM is trained, we use it to predict the labels of test-set apps. Subsequent to predicting an app $\textbf{x}$ to be malicious, \tool{} performs the following:
\begin{itemize}[leftmargin=*]
\setlength\itemsep{0em}
\item Awards an m-score to every node in $\textbf{x}$'s CICFG. This helps to locate basic blocks that perform malice operations. We choose to locate malice nodes from CICFG as all the five PRGs used in \tool{} are its abstractions and hence all contextual subgraph features emerging from individual views could be traced back to CICFG, only.
\item The m-scores of all the methods and classes in \textbf{x} are deduced by aggregating the m-scores of their constituent basic blocks. This helps to locate larger portions of malice codes capable enough to explain \textbf{x}'s attacks.
\end{itemize}

Our procedure to award m-scores requires interpreting the predictions of MKL-SVM. Before delving into the details of  m-score computation algorithm, we introduce some preliminaries required for this interpretation.

\textbf{Interpretability: primal vs. dual formulations.} The formulation of SVM discussed in eqs. (4) to (6) is known as '\textit{dual formulation}' and one mentioned in eq. (8) and (9) is known '\textit{primal formulation}'. In general, SVMs built in the former formulation are uninterpretable. 
On the contrary, interpreting the predictions in primal formulation is a well-studied problem as discussed in \cite{Drebin,Adagio,lime}. These methods enable us to determine the contribution of each feature to the final class prediction as described below.

Based on eq. (8), for a given sample $\textbf{x}$, during the prediction of its label, $f(\textbf{x})$, the contributions of individual features towards placing the sample on the positive (or negative) side of the decision boundary are identified by performing a point-wise multiplication of the weight and sample's vectors, i.e., the contribution of feature $\mathfrak{f}$ is deduced as: 
\begin{equation}
c^\mathfrak{f} = w^\mathfrak{f} \cdot x^\mathfrak{f}
\end{equation}
where $x^\mathfrak{f}$ denotes the frequency of occurrence of feature $\mathfrak{f}$ in $\textbf{x}$ and $ w^\mathfrak{f} $ is its relative importance. Meaning, high frequency of features with high positive (or negative) weight would result in large positive (or negative) value of $c^\mathfrak{f}$, pushing sample $\textbf{x}$ significantly towards the positive (or negative) side of the decision boundary.%\footnote{Understandably, $\langle \overrightarrow{\textbf{w}},\overrightarrow{\textbf{x}} \rangle = \sum_{\mathfrak{f} = 1}^{|\overrightarrow{\textbf{x}}|} c^\mathfrak{f}$ and therefore, $f(\textbf{x}) = sign \big( \sum_{\mathfrak{f} = 1}^{|\overrightarrow{\textbf{x}}|} c^\mathfrak{f} \big)$.}. 

In the malware detection case, features with large positive and negative values of $c^\mathfrak{f}$ characterize strong malice and benign behaviors, respectively. 
%{\color{red}In fact, \textsc{Drebin} \cite{Drebin}, a well-known interpretable malware detection solution ranked the top features with large positive $c^\mathfrak{f}$ values and demonstrated that they represent the malice actions performed in the sample.} 
This interpretablity procedure could be adopted to locate contextual subgraph features from our PRGs that contribute significantly to placing a sample on the positive (i.e., malicious) side of the decision boundary.

\textbf{Interpretability in MKL scenario.} However, adopting this procedure with MKL-SVMs is not straight-forward, as they are learnt strictly in the dual formulation. More specifically, we need the following in the MKL scenario to offer interpretations:
\begin{itemize}[leftmargin=*]
\setlength\itemsep{0em}
	\item An explicit and composite representation of a sample $ \textbf{x} $ with features from all the base kernels.
	
	\item A weight vector which quantifies the weights of features from all the base kernels. %incorporating the relative importances of both the feature and the kernel.
\end{itemize} 
These two vectors are not obtainable with SMO-MKL. Hence, we obtain the primal version of our MKL-SVM and use it for interpreting its predictions and computing the m-scores as explained in algorithm \ref{algo:predict}.

\begin{algorithm}[t]
	\small
	\caption{\textsc{PredictAndInterpret}}
	\label{algo:predict}
	\SetKwInOut{Input}{input}
	\SetKwInOut{Output}{output}
	\SetKwInOut{Given}{given}
	\Input
	{$\textbf{x}_{train} $ | Training set samples. \newline
		$y_{train}$ | Training set labels. \newline
		$\alpha$ | Support vectors learnt through linear MKL in the dual formulation. \newline
		$\beta = \{\beta_{v_1}, \beta_{v_2}, ...\}$ | Weights of base kernels learnt through linear MKL. \newline
		$\textbf{x}_{test} $ | Test-set samples. \newline
		$ CICFG_{test} $ | CICFGs of test-set samples.
	}
	\Output
	{ $\hat{y}_{test}$ | Predicted labels of the test-set samples.\newline
		m-scores\textsubscript{test} | Maliciousness scores CICFG nodes of test-set samples.
	}
	
	\Begin{
		Initialize: $\hat{y}_{test} = \{\}$ and m-scores\textsubscript{test}$ = \{\}$\\
		\tcp{compute composite representations of training-set samples}
		\For{$ i \in [1,|\textbf{x}_{train}|] $ }
		{
			$\overrightarrow{\mathsf{\textbf{X}}}^{(i)}  = \bigoplus_{v \in V} \sqrt{\beta_v} \cdot  \overrightarrow{\textbf{x}}_v^{(i)}$
			% = [\mathsf{x}^1,\mathsf{x}^2,...]^T $\\
		}
		\tcp{compute the composite weight vector}
		$\overrightarrow{\mathsf{\textbf{W}}}= \sum_{i=1}^{|\textbf{x}_{train}|} \alpha^{(i)}y^{(i)}\overrightarrow{\mathsf{\textbf{X}}}^{(i)}	= [\mathsf{w}^1, \mathsf{w}^2, ... ]^T$

		\For{$\textbf{x} \in x_{test}$}
		{
			\tcp{compute \textbf{x}'s composite representation}
			$\overrightarrow{\mathsf{\textbf{X}}}  = \bigoplus_{v \in V} \sqrt{\beta_v} \cdot  \overrightarrow{\textbf{x}}_v = [\mathsf{x}^1, \mathsf{x}^2, ... ]^T$\\
			\tcp{Predict \textbf{x}'s label in the primal formulation}
			$ f(\textbf{x}) $ = $Sign$ ($ \langle \overrightarrow{\mathsf{\textbf{X}}}, \overrightarrow{\mathsf{\textbf{W}}} \rangle$) = $sign \big( \sum_{\mathfrak{f}=1}^{|\overrightarrow{\mathsf{\textbf{X}}}|} \mathsf{w}^\mathfrak{f}\mathsf{x}^\mathfrak{f} \big)$\\
			$\hat{y}_{test}[\textbf{x}] = f(\textbf{x})$\\
			
			\If{$f(\textbf{x})$ is $+1 $} 
			{
				m-scores\textsubscript{test}[\textbf{x}] = \textsc{AwardMaliciousnessScore} ($\overrightarrow{\mathsf{\textbf{X}}}, \ CICFG_{test}[\textbf{x}],\ ,\overrightarrow{\mathsf{\textbf{W}}}$)
			}
		}

		\textbf{return} $\hat{y}_{test}$, m-scores\textsubscript{test}
	}
\end{algorithm}

\textbf{Algorithm: \textsc{PredictAndInterpret}.} 
The algorithm takes as inputs the training samples ($x_{train}$) along with their labels ($y_{train}$), the support vectors ($\alpha$) and base kernel weights ($\beta$) learnt in the dual formulation, test samples ($x_{test}$) and their CIFCGs ($CICFG_{test}$) for which m-scores have be computed.

Firstly, The predicted labels of test samples ($\hat{y}_{test}$) and their m-scored CICFGs (m-scores\textsubscript{test}) are initialized to empty sets (line 2).

Then the composite representation of every training sample $\overrightarrow{\mathsf{\textbf{X}}}^{(i)}$ is computed  in lines 3 and 4. Here, $\oplus$ denotes concatenation operation and $\overrightarrow{\mathsf{\textbf{X}}} $ is obtained by concatenating the feature vectors from individual base kernels after scaling them with corresponding base kernel weights. It could be seen that,
\begin{align*}
\langle \overrightarrow{\mathsf{\textbf{X}}}^{(i)},\overrightarrow{\mathsf{\textbf{X}}}^{(j)} \rangle  = \sum_{v \in V} \beta_v \langle \overrightarrow{\textbf{x}}^{(i)}_v,\overrightarrow{\textbf{x}}^{(j)}_v \rangle = \sum_{v \in V} \beta_v k_v(\textbf{x}^{(i)},\textbf{x}^{(j)}) = \\ k_{comb} (\textbf{x}^{(i)},\textbf{x}^{(j)})
\end{align*}
Thus, the composite representations encompass features from all views incorporating their relative importances.
Subsequently, the composite weight vector, $\overrightarrow{\mathsf{\textbf{W}}} $, required for predictions and interpretations is obtained from representations of support vector instances in line 5. Clearly, in this vector, $\mathsf{w}^\mathfrak{f}$, the weight of the base kernel feature $\mathfrak{f}$ accounts for both the kernel's and feature's relative importance. 
 
Once the weight vector of the MKL-SVM is obtained, we predict the label of every test-set sample using the primal formulation\footnote{From eq. (9), it could be noted that, the prediction made in this fashion will be equivalent to one made with a linear SVM learnt as an optimization on $\overrightarrow{\mathsf{\textbf{W}}}$ as follows:$	\min_{\overrightarrow{\mathsf{\textbf{W}}}} ||\overrightarrow{\mathsf{\textbf{W}}}||^2 + \sum_{i=1}^N  max(0,1-y^{(i)}f(\overrightarrow{\mathsf{\textbf{X}}}^{(i)}))$.} in line 8. 
If a sample, \textbf{x}, is predicted to be malware (i.e., $f(\textbf{x})\  is\ +1$), we compute the m-scores of all nodes in its CICFG in line 11. The detailed procedure for m-score computation is explained separately in algorithm \ref{algo:m-score}. Once all the test-set samples are subjected to prediction and m-score computation the results are returned in line 12.

\begin{algorithm}[t]
	\small
	\caption{\textsc{AwardMaliciousnessScore} }
	\label{algo:m-score}
	\SetKwInOut{Input}{input}
	\SetKwInOut{Output}{output}
	\SetKwInOut{Given}{given}
	\Input
	{$\overrightarrow{\mathsf{\textbf{X}}}$ | Composite representation of test-set sample \textbf{x} predicted to be malware. \newline
		$ CICFG = (N_i,E_i) $ | \textbf{x}'s CICFG. \newline
		$\overrightarrow{\mathsf{\textbf{W}}}$ | Composite weight vector. 
	}
	\Output
	{ $\text{m-scores}$ - maliciousness scores of all the nodes in \textbf{x}'s CICFG.
	}
	
	\Begin{
		
		Initialize: $\text{m-scores} = \{\}$\\
		
		\tcp{Calculate maliciousness scores for every node in \textbf{x}'s CICFG}
		\For{$n \in N_i$} 
		{
			\tcp{Compute vector representation of node \textit{n}}
			$\overrightarrow{\mathsf{\textbf{X}}}_n = I_n (\mathfrak{f},n) \cdot \overrightarrow{\mathsf{\textbf{X}}} = [\mathsf{x}^1_n, \mathsf{x}^2_n, ... ]^T$\\
			\tcp{ $ I_n(\mathfrak{f},n) =  
				\begin{cases}
				1,& \text{if feature $ \mathfrak{f} $ emerges from node \textit{n}}\\
				0,& \text{otherwise}
				\end{cases} $}
			\tcp{Arrive at maliciousness score of node \textit{n}}
			$ \text{m-scores} [n] = \langle \overrightarrow{\mathsf{\textbf{W}}}, \overrightarrow{\mathsf{\textbf{X}}}_n \rangle =  \sum_{\mathfrak{f}=1}^{|\overrightarrow{\mathsf{\textbf{X}}}_n|} \mathsf{w}^\mathfrak{f}\mathsf{x}^\mathfrak{f}_n 
			= \sum_{\mathfrak{f}=1}^{|\overrightarrow{\mathsf{\textbf{X}}}_n|} \mathsf{c}^\mathfrak{f}_n$
		}

		\textbf{return} $\text{m-scores} $
	}
\end{algorithm}

\textbf{Algorithm: \textsc{AwardMaliciousnessScore}.}
Given the CICFG of a sample $\textbf{x}$ that is predicted to be a malware, the goal of this algorithm is to award m-scores to each of its nodes that quantify the severity of their malice operations. To achieve this, the algorithm needs the explicit composite representation of \textbf{x} (i.e., $\overrightarrow{\mathsf{\textbf{X}}}$) and the MKL-SVM's weight vector (i.e., $\overrightarrow{\mathsf{\textbf{W}}}$). These values are passed as its inputs (from line 11 of algorithm \ref{algo:predict}).

To begin with, a dictionary of m-scores of all the nodes are initialized (line 2). Subsequently, we loop through every node $n \in N_i$ in \textbf{x}'s CICFG and compute their m-scores with a 2-step procedure in lines 4 and 5. %In the first step we compute the vectorial representation of a CICFG node $n$ (line 4) and in the second step, we leverage on this representation to arrive at $ n $'s m-score (line 5).

In line 4, node $n$'s composite representation, $\overrightarrow{\mathsf{\textbf{X}}}_n$, is obtained from the sample's vector by unmasking only the contextual subgraph features that emerge from $n$. The identifier function $I_n$ helps this unmasking. Clearly, $\overrightarrow{\mathsf{\textbf{X}}}_n$ encompasses features from all the five views. 

Once node $n$'s representation is arrived at, we could calculate the contributions of individual features emerging from $n$ to the final prediction of $\overrightarrow{\mathsf{\textbf{X}}}$ using eq. (12). That is, the contribution of a feature $\mathfrak{f}$ from node $n$ to the final prediction $ f(\textbf{x}) $ is: $\mathsf{c}^\mathfrak{f}_n = \mathsf{w}^\mathfrak{f} \cdot \mathsf{x}^\mathfrak{f}_n$ (where $ \mathsf{x}^\mathfrak{f}_n $ denotes the frequency of occurrence of $\mathfrak{f}$ in $n$).
Therefore, to calculate the m-score of $n$, we just need to aggregate $\mathsf{c}^\mathfrak{f}_n$ as in line 5.
Finally, the m-scores of all nodes in \textbf{x}'s CICFG thus computed are returned in line 6.

Once the testing and interpretation phase finishes, we would have the predictions for all test-set samples ($\hat{y}_{test}$) and the CICFGs of predicted malware with m-scores of their nodes (m-scores\textsubscript{test}). Subsequently, m-scores of nodes are aggregated to compute the same for classes and methods encompassing them as follows:
\begin{equation}
\text{m-score}(m) = \sum_{n \in N_i} I_m(n,m) \cdot \text{m-score}(n)
\end{equation}
\begin{equation}
\text{m-score}(c) = \sum_{n \in N_i} I_c(n,c) \cdot \text{m-score}(n)
\end{equation}
where the indicator function $I_m$ and $I_c$ are defined as,

{\small \[
I_m(n,m) =  
\begin{cases}
1,& \text{if basic block $ n $ is contained in method $ m $}\\
0,& \text{otherwise}
\end{cases}
\]}

{\small \[
I_c(n,c) =  
\begin{cases}
1,& \text{if basic block $ n $ is contained in class $ c $}\\
0,& \text{otherwise}
\end{cases}
\]}

Computing the m-scores of methods and classes culminates \tool's automated detection and malicious code localization procedures. Subsequently, analysts could investigate methods and classes with high scores so as to understand malware's attacks and evasion footprints. 

It could be easily seen that this process of computing m-scores could be used with base kernels to determine scores from individual views (i.e., {\tt API} kernel m-scores will depend only on CADG contextual subgraphs and so on.). However, thanks to the multi-view analysis, the MKL based m-scores are more comprehensive and robust in locating malice code than those from individual views (demonstrated later through evaluations in \S \ref{ss:lme}).

In sum, for interpretable multi-view detection, we have trained the MKL-SVM in the dual formulation and predicted the labels of the test-set apps in the primal formulation, which helps to compute the significance of every feature towards the final prediction. To the best of our knowledge, there is no other work that switches MKL-SVMs formulations like ours, as this interpretablity requirement is unique to our goal of malicious code localization.

%\textbf{Advantages gained through MKL.} 
Overall, \tool{} reaps the following advantages through its MKL: 
\begin{enumerate} [label=(\roman*)]
	\item MKL elegantly combines features from five different views of the app, in a way which allows the learning algorithm to take advantage of all of them simultaneously.
	
	\item  \tool’s learning is extendable in the sense that new semantic views (e.g., dynamic analysis or data-flows based views) could be easily added to the model without complicating the final result. 
	
	\item \tool’s detection process is parallelizable: constructing representations and computing kernel values for unseen testset apps can all be done in parallel, the implication being that larger datasets can easily be handled.
	
	\item Interpretability achieved over MKL allows \tool{} to perform precise multi-view malicious code localization.
\end{enumerate}

\section{Experimental Design and Implementation}
\label{sec:edi}
We conducted several large-scale experiments to evaluate \tool's accuracy, efficiency and malicious code localization capabilities. We also perform comparative analysis with three state-of-the-art Android malware detection solutions. In this section, experimental design aspects such as research questions (RQs) addressed, datasets used, evaluation setup and metrics are presented along
with implementation details.

\subsection{Research Questions}
\label{subsec:exp_des}

We intend to address the following RQs through our evaluations:

\textbf{(RQ1 Accuracy)} How accurate are \tool's individual views in detecting malware and how does it benefit from appropriately combining them?

Accuracy under different experimental settings are investigated through the following sub-RQs:

(RQ1.1) How accurate is \tool{} in detecting unseen malware when trained with an up-to-date dataset and how does it compare to \soa{} approaches?

(RQ1.2) How accurate is \tool{} in detecting unseen malware when trained with a dataset that is historically anterior to the evaluation set?

(RQ1.3) How accurate is \tool{} in detecting recent malware apps collected in-the-wild?

(RQ1.4) Which views of \tool{} are most (and least) effective and does combining them through MKL offer significant improvements?

\textbf{(RQ2 Efficiency)} How efficient are \tool's individual views in terms of overall training and prediction time and does combining them incur significant overhead?

\textbf{(RQ3 Locating malice)} How accurately does \tool{} locate malice code in a given sample and does it explain the malicious behavior exhibited by the sample?

\subsection{Datasets \& Experiments}
We conducted experiments with both benchmark datasets and apps collected in-the-wild. These datasets with  details such as number of samples and time of compilation are presented in Table \ref{tab:datasets}. A total of 60,561 apps have been used in our evaluations. The design of all our experiments are summarized in Table \ref{tab:ed}.

\begin{table}[t]
	\centering
	\scriptsize
	\setlength\tabcolsep{5pt}
	\caption{Datasets used in evaluation}
	\label{tab:datasets}
	\begin{tabular}{|C{4cm}|C{4cm}|C{2cm}|C{2.75cm}|}
		\hline
		{\textbf{Dataset Category}}                    & {\bf Dataset Name}       & {\bf \# of samples} & {\bf \begin{tabular}[c]{@{}c@{}}Time of \\ compilation\end{tabular}} \\ \hline \hline
		\multirow{2}{*}{\textbf{Malware Datasets}}   & \textsc{Drebin} (DR) \cite{Drebin}       & 5,560             & Aug'10 - Oct'12                                                        \\  
		& Virus-share (VS) \cite{VS} & 24,317            & May'13 - Mar'14                                                        \\ \hline
		\multirow{2}{*}{\textbf{Benign Datasets}}     & Google Play 1 (GP1)      & 5,000             & \multirow{2}{*}{Jul'12 - May'14}                                              \\ 
		& Google Play 2 (GP2)      & 10,000            &                                                                      \\ \hline
		\multirow{5}{*}{\textbf{Wild Datasets}} & AndroidDrawer (AD)  \cite{AndDr}     & 2,399             & \multirow{5}{*}{\color{black}{Aug'13 - Sep'16}}                                       \\ 
		& AnZhi (AZ)   \cite{AnZhi}            & 3,027             &                                                                      \\ 
		& AppsApk (AA)   \cite{AppsApk}          & 2,481             &                                                                      \\ 
		& FDroid (FD) \cite{FDroid}             & 1,007             &                                                                      \\ 
		& SlideMe (SM) \cite{SlideMe}            & 5,770             &                                                                      \\ \hline
		{\bf Dataset annotated with locations of malice code} & \textsc{Mystique} (MYST) \cite{mist} & 3,000 & Dec'15 \\ \hline
	\end{tabular}
\end{table}

\begin{table*}[ht!]
	\centering
	\scriptsize
	\setlength\tabcolsep{0pt}
	\caption{List of experiments conducted and the composition of training and test-sets}
	\label{tab:ed}
	\begin{tabular}{|C{2.5cm}|C{1.6cm}|C{1.5cm}|C{1.5cm}|C{3.85cm}|C{4cm}|}
		\hline
		{\bf \begin{tabular}[c]{@{}c@{}}Experiment\\ type\end{tabular}}                         & {\bf \begin{tabular}[c]{@{}c@{}}Experiment\\ \#\end{tabular}} & {\bf Malware Samples} & {\bf Benign Samples} & {\bf Training set}                                                                     & {\bf Test-set}                                                       \\ \hline \hline
		\multirow{3}{*}{{\bf \begin{tabular}[c]{@{}c@{}}Controlled \\ Experiment\end{tabular}}} & {\tt CE1}                                                           & DR             & GP1                  & 70\% of samples                                                     & 30\% of samples                                          \\ \cline{2-6} 
		& {\tt CE2}                                                           & VS                 & GP2                  & 70\% of samples                                                     & 30\% of samples                                          \\ \cline{2-6} 
		& {\tt CE3}                                                           & DR, VS     & GP1, GP2            & \begin{tabular}[c]{@{}c@{}}Malware: \textsc{Drebin}\\ Benign: GP1\end{tabular}                  & \begin{tabular}[c]{@{}c@{}}Malware: VS\\ Benign: GP2\end{tabular} \\ \hline
		{\bf Wild Experiment}                                                                       & {\tt WEx}                                                           & DR, VS     & GP1, GP2            & \begin{tabular}[c]{@{}c@{}}Malware: DR + VS\\ Benign: GP1 + GP2\end{tabular} & \begin{tabular}[c]{@{}c@{}}Third-party market apps\\ (AD + AZ + AA + FD + SM)\end{tabular}   \\ \hline
		{\bf Malicious code localization }                                                                       & {\tt MCLEx}                                                           & MYST                 & GP1                 & {2/3}\textsuperscript{rd} of samples                                                     & {1/3}\textsuperscript{rd} of samples                                         \\ \hline
	\end{tabular}
\end{table*}

\subsubsection{Controlled Experiments}
Controlled experiments were conducted on malware samples from well-known benchmark datasets and benign apps from Google Play. Three controlled experiments ({\tt CE1}, {\tt CE2} and {\tt CE3}) were conducted as described below.\\
\textbf{{\tt CE1}:} 5,560 malware apps from DR and 5,000 benign apps from GP1 collections were used to form the dataset for experiment {\tt CE1}. The model is trained using 70\% of these samples chosen at random and is tested for accuracy on the remaining 30\%  samples.% The standard 5-fold cross-validation (CV) procedure is used for model selection.
\\
\textbf{{\tt CE2}:} 24,317 malware apps from VS and 10,000 benign apps from GP2 collections were used to form the
dataset for experiment {\tt CE2}. The training and test-set apps ratio
% and the CV procedure 
is same as {\tt CE1}.\\
\textbf{{\tt CE3}:} It could be observed that the process followed in {\tt CE1} and {\tt CE2} (i.e., splitting the malware and benign samples randomly into training and test-sets and performing evaluation) is followed in almost all the previous malware detection methods such as \cite{Adagio,Drebin,Mudflow,AppContext}. However, this type of evaluation has two issues:\\
(1)	Malware in benchmark datasets were collected at a particular point in time and hence are homogeneous in terms of their attack vectors. However, malware continue to evolve and more sophisticated variants are produced subsequent to publishing such datasets \cite{CSBD,mama}.\\
(2) As observed by Allix et al. \cite{HM}, in these experiments, \textit{samples in the training set may be historically posterior to those in the test-set}. While, in the real-world/AV industry settings, when a new unseen app must be processed for detection, the training sets used are, necessarily, historically anterior to the new app. This constraint is not considered in experiments similar to {\tt CE1} and {\tt CE2}.

We address these two issue, in experiment {\tt CE3}, by enforcing that the training set used for building the classifier is historically anterior to the test-set. We achieve this by using the samples from DR dataset which were collected from 2010 to 2012 to train the detection model (along with GP1 benign apps) and we use the samples from VS dataset which were collected from 2013 to 2014 to test the classifier (along with GP2 benign apps). 

\subsubsection{Wild Experiments}
The controlled experiments {\tt CE1}, {\tt CE2} and {\tt CE3} were conducted on malware from benchmark datasets.
A common observation is that real-world malware, due to their rapid evolution are more challenging to detect that the ones in the benchmark datasets \cite{mist}.
A technique's effectiveness in detecting malware in the wild could not be determined through testing on such outdated homogeneous datasets.
To address this, we also test our model on a large collection of recent apps from popular third-party markets (in experiment {\tt WEx}).
To this end, a total of 14,684 apps from five different third-party markets were collected from Aug. 2013 to Sep. 2016. To test the model on these apps, we need the ground truth labels of these apps (i.e., whether they are malicious or benign). 
To this end, following the software security research practices proposed in \cite{Drebin} and \cite{CSBD}, we leveraged on the VirusTotal web portal\footnote{\url{https://www.virustotal.com}} to infer their ground truth labels. Out of these apps, 6,128 are found to be malware. %Hence, this dataset is also roughly balanced in terms of the proportion of malware and benign apps.
Thus, in {\tt WEx}, all the malware apps from DR and VS (29,877 in total) and benign apps from GP1 and GP2 (15,000 in total) collections were used to train the classifier. The test-set comprises of 14,684 wild apps.

\subsubsection{Malicious Code Localization Experiment}
In this experiment, we intend to evaluate \tool's capabilities to locate the malicious code in a given sample. More specifically, we explore whether it could locate malice methods or classes involved in the sample's attacks.

\textbf{Quantitative Evaluation.}
As mentioned earlier, though a vast body of literature on Android malware detection approaches exists, none of them systematically addressed the problem of locating malice code in given sample. This is partly due to the fact that  none of existing datasets (incl. DR \cite{Drebin}, VS \cite{grafting}, AndroZoo \cite{androzoo}) provide ground-truth on the location of malice code\footnote{Recently, Li et al. \cite{grafting} provided a dataset of repackaged apps of the form: $(app1, app2)$, where $app1$ is the original (benign) app and $(app2)$ is the repackaged version of $app1$. However, they do not ascertain whether or not the new code injected in $app2$ is malicious. In fact, exploring this dataset, we observe that a majority of the repackaged apps were \textit{adware} or other type of PHAs. Hence, we refrain from using this dataset which lacks precise ground truth labels on malice methods and classes in our experiments.} such as names of methods/classes involving in malice operations. They just provide labels to ascertain whether a sample holistically is benign or malicious. 
Obviously, with these datasets, one could not quantitatively evaluate malicious code localization capabilities. In order to address this, we extended an existing dataset as follows. Recently, \textsc{Mystique} \cite{mist} proposed an Evolutionary Computation based method to automatically generate new malware samples learning from attack and evasion strategies of benchmark malware. \textsc{Mystique} provided a dataset of 10,000 such automatically generated malware with the names of the classes that contain malice code. However, almost all the code in these samples are either malicious or from commonly used libraries (e.g., \textit{android.support}) and there are no benign functionalities. In other words, none of these apps are repackaged malware and hence do not cater well to the real-world needs of locating malice code in repackaged malware. Hence to extend it by randomly choosing 3,000 apps from this dataset and piggybacked the same on benign apps from Google Play. Hence, for each of these apps we are certain of the following: (i) they contain both benign and malice code, (ii) names of the classes that contain malice code.
We refer to this dataset as MYST and use the same in experiment {\tt MCLEx}, where we investigate \tool's malicious code localization capabilities quantitatively, as follows. 

We train our model using 2,000 malware from MYST dataset and 2,000 benign apps (that were not used for piggybacking). Subsequently, we test the model on the remaining 1,000 MYST samples. Remember, during testing, \tool{} assigns m-scores to every class in an app. Hence, we investigate whether the classes with highest m-scores are indeed malicious using MYST's ground-truth.

\textbf{Qualitative Evaluation.}
To perform qualitative evaluation we do not need manual annotations on classes/methods containing malice code. All we need are samples that contain malice code at least in one of their classes/methods and we could use \tool{} on them to investigate if it locates such code. Hence, for qualitative evaluation, we choose the DR dataset\footnote{More than 80\% of samples in this dataset are piggybacked malware thus making this dataset amenable for our qualitative analysis \cite{grafting}.} that contains real-world malware with malice code spread across many methods and classes. The experimental settings in {\tt CE1} are reused in this evaluation. We manually investigate whether classes and methods with high m-scores correspond to the sample's attacks.

\subsection{Experimental Setup}
All the experiments were conducted on a server with 20 cores of Intel(R) Xeon(R) CPU E5-2699 v3 @ 2.30GHz and 200 GB RAM running Ubuntu 14.04.

\subsection{Implementation and Comparative Analysis}
\label{ss:impl}

\tool{} is implemented in approximately 9,700 lines of Python, Java and C++ code. Androguard \cite{AG} and Soot \cite{Soot} have been used to build the PRGs and infer the reachability contexts of PRG nodes. For SMO-MKL functionalities, the source code provided by SVN Vishwanathan et al. \cite{SMO-MKL} has been used. %For locating malice through interpretation, SVM from Scikit-learn\footnote{\url{http://scikit-learn.org/}} has been used.

\textbf{Comparison with state-of-the-art solutions.}
Our approach is compared against three \soa{} ML based Android malware detection solutions, namely, \textsc{Drebin} \cite{Drebin}, Allix et al. \cite{CSBD} and \textsc{Adagio}\cite {Adagio}. To this end, we re-implemented \textsc{Drebin} and Allix et al.’s approaches. For \textsc{Adagio}, an open-source implementation \cite{adagiocode} provided by the authors is used.
Since the accuracy of these solutions predominantly depend on the features they use, we briefly introduce them here. 

\textbf{Drebin \cite{Drebin}} is well-known for its scalable and explainable detection. It extracts light-weight semantic features such as APIs and permissions used, URLs accessed, names of components from apps and subsequently, trains a linear SVM to distinguish malware from benign apps.

\textbf{Allix et al. \cite{CSBD}} proposed another scalable approach using signatures of basis blocks in CFGs. Therefore, we refer to this technique as CFG Signature Based Detection (CSBD) in the reminder of the paper. CSBD constructs CFGs of individual methods and encodes them as text-signatures following  Cesare and Xiang's grammar \cite{CFGGrammar}. Subsequently, a RF classifier
is trained with these signatures to detect malware subsequently used as features. 

\textbf{Adagio \cite{Adagio}} constructs CGs and uses byte-code instructions to assign labels to nodes. NHGK \cite{NHGK} is used to extract CG neighborhoods as features and a histogram-intersection (HI) kernel SVM is trained to detect malware. \textsc{Adagio} uses HI kernel in the primal formulation to achieve interpretable results. %This causes severe scalability issues. %To circumvent this we have extended \textsc{Adagio}'s code by using HI kernels dual formulation and understandably this improved the efficiency significantly.

We re-implemented \textsc{Drebin} and CSBD in 1400 and 900 lines (approx.) of Python code, respectively. Authenticity and correctness of our re-implementations is verified as we observe their accuracy and scalability values very similar to the ones reported in the original work on similar experiments (see \S \ref{sec:rd}). Besides this, re-implementations have been done in consultation with the authors of original work.

%\vspace{-5mm}
\subsection{Evaluation metrics}
Standard evaluation metrics such as Precision, Recall and F-measure are used to determine the effectiveness of malware detection. All these values are in the range [0, 1]. Higher values indicate accurate detection. Efficiency is determined in terms of training and testing durations (in seconds). Lower training and testing durations indicate scalable detection. For evaluating malicious code localization, False Positive Rate (FPR) and False Negative Rate (FNR) measures are used. These are expressed as percentage values. Lower FPR and FNR indicate precision and completeness in detection, respectively.  

\section{Results and Discussions}
\label{sec:rd}
The evaluation, results and relevant discussions pertaining to each of the RQs is presented in this section.
The accuracy and efficiency results for controlled and in-the-wild experiments is presented in subsections \S \ref{ss:acc} and \S \ref{ss:eff}, respectively. For malicious code localization, a qualitative evaluation which involves case studies on two well-known malware families and a quantitative evaluation on the homegrown dataset are presented in \S \ref{ss:lme}.

\subsection {RQ1: Accuracy}
\label {ss:acc}
\subsubsection{RQ1.1 Accuracy on benchmark datasets}

As mentioned earlier, in experiments {\tt CE1} and {\tt CE2}, 70\% of samples were randomly chosen from the evaluation datasets and used for training the classifier and the remaining 30\% samples are used to test its performance. The hyper-parameters of classifier are determined on the training set (using 5-fold cross-validation), whereas the test-set is only used for determining the final detection performance. We repeat this procedure five times and average the results. 
In order to study the effectiveness of individual views, we report the prediction results using individual base kernels and the uniform kernel (which is the mean of all base kernels). 

The results for experiments {\tt CE1} and {\tt CE2} along with comparison to \soa{} techniques is presented in tables \ref{tab:ce1} and \ref{tab:ce2}, respectively. The following inferences are drawn from these tables:

\begin{table}[t]
	\centering
	\scriptsize
	\captionsetup{justification=centering}
	\caption{{\tt CE1} Results: Precision, Recall \& F-measure (avg $\pm$ std.)}
	\label{tab:ce1}
	\begin{tabular}{|l|l|l|l|}
		\hline
		Technique        & P & R & F \\ \hline \hline
		{{\tt API} kernel} & 0.974 ($\pm$0.004)  & 0.989 ($\pm$0.001) & 0.982 ($\pm$0.004)  \\ 
		{{\tt Permission} kernel} & 0.960 ($\pm$0.010)  & 0.978 ($\pm$0.008) & 0.970 ($\pm$0.011)
		\\ 
		{{\tt Src-sink} kernel} & 0.802 ($\pm$0.047)  & 0.734 ($\pm$0.012) & 0.766 ($\pm$0.008)
		\\ 
		{{\tt Instruction} kernel} & 0.967 ($\pm$0.005)  & 0.989 ($\pm$0.004) & 0.979 ($\pm$0.004) 
		\\ 
		{{\tt Signature} kernel} & 0.971 ($\pm$0.005)  & \textbf{0.990} ($\pm$0.008) & 0.981 ($\pm$0.005)
		\\ \hline
		{Uniform kernel} & 0.987 ($\pm$0.003)  & 0.982 ($\pm$0.002) & \textbf{0.985} ($\pm$0.003) 
		\\ \hline
		{\tool} & \textbf{0.988} ($\pm$0.004)  & 0.982 ($\pm$0.004) & \textbf{0.985} ($\pm$0.003)
		\\ \hline
		{Drebin}\cite{Drebin}      & 0.980 ($\pm$0.002) & \textbf{0.990}($\pm$0.001) & \textbf{0.985} ($\pm$0.002)
		\\ \hline
		{Adagio}\cite{Adagio} & 0.960 & 0.974 & 0.967
		\\ \hline
		{CSBD}\cite{CSBD} & 0.954 ($\pm$0.005) & 0.987 ($\pm$0.008) & 0.970 ($\pm$0.005)  \\ \hline
		
	\end{tabular}
\end{table}

\begin{table}[t]
	\captionsetup{justification=centering}
	\centering
	\scriptsize
	\caption{{\tt CE2} Results: Precision, Recall \& F-measure (avg $\pm$ std.)}

	\label{tab:ce2}
	\begin{tabular}{|l|l|l|l|}
		\hline
		Technique        & P & R & F \\ \hline \hline
		{{\tt API} kernel} & \textbf{0.966} ($\pm$0.010) & 0.943 ($\pm$0.016) & 0.954 ($\pm$0.024) \\ 
		
		{{\tt Permission} kernel}  & 0.962 ($\pm$0.022) & 0.931 ($\pm$0.038) & 0.946 ($\pm$0.017) \\
		
		{{\tt Src-sink} kernel} & 0.710 ($\pm$0.035) & 0.728 ($\pm$0.052) & 0.719 ($\pm$0.044) \\ 
		
		{{\tt Instruction} kernel} & 0.954 ($\pm$0.010) & 0.946 ($\pm$0.018) & 0.950 ($\pm$0.010) \\ 
		{{\tt Signature} kernel} & 0.951 ($\pm$0.014) & 0.944 ($\pm$0.018) & 0.948 ($\pm$0.011) \\ \hline
		{Uniform kernel} & 0.964 ($\pm$0.013) & 0.952 ($\pm$0.009) & 0.958 ($\pm$0.012) \\ \hline
		{\tool} & \textbf{0.966} ($\pm$0.015) & \textbf{0.980} ($\pm$0.008) & \textbf{0.973} ($\pm$0.009) \\ \hline
		
		{Drebin}\cite{Drebin}      & 0.960 ($\pm$0.007) & \textbf{0.980} ($\pm$0.009)  & 0.971 ($\pm$0.006) \\ \hline
		{Adagio}\cite{Adagio}      & 0.960 & 0.960 & 0.960 \\ \hline
		{CSBD}\cite{CSBD}     & 0.914 ($\pm$0.028) & 0.958 ($\pm$0.008) & 0.938 ($\pm$0.015) \\ \hline
		
	\end{tabular}
\end{table}

\begin{itemize} [leftmargin=*]
	\setlength\itemsep{0em}
	\item At the outset, we observe that all individual views have certain effectiveness in detecting malware. This is reflected by the fact that all the base kernels get more than 75\% F-measure in both the experiments. In fact, 4 out of 5 base kernels offer comparable F-measures to \soa{} approaches.
	\item Out of the base kernels, {\tt API} kernel achieves the best performance in both the experiments. Meaning, context-aware structural API dependencies turn out to be excellent features for detecting malware. In fact, this observation goes hand-in-hand with the fact that API related features (API frequencies and ngrams \cite{Drebin,droidapiminer}, API sequences \cite{DroidMiner}, subgraphs \cite{AppContext,MLMalDetect,droidol,cwlk,sg2vec}, dependencies \cite{Mudflow,reveal}, etc.) are the most popular features in Android malware detection literature.
	
	\item The performances of instruction and signature kernels are very close in both the experiments and are marginally less effective than those of {\tt API} kernel. This is because both these kernels capture structural dependencies from the apps that are sparse, obfuscation resilient and abstract. 
	
	\item {\tt Permission} kernel achieves significantly lesser standalone F-measure than the three above mentioned kernels. Understandably, permissions are more coarse-grained features compared to API and instruction sequences. This leads to many misclassifications thereby hampering the kernel's accuracy. 
	
	\item {\tt Src-sink} kernel obtains the least F-measure indicating that it exhibits least malware detection potential. Note that CSSDG actually captures control flows across source and sink nodes and this does not necessarily mean there are data-flows (i.e., information leak) in the suspected control flow paths\footnote{Remember, we intend to avoid computing expensive data-flows in the app and believe other views (computed at much lesser expense) would complement and mitigate the absence of data-flow related features.}. We conjecture that considering control-flow paths as proxies for detecting information leak attacks leads to considerable false positives and thus resulting in a poor detection.  
	
	\item The main justifications for using MKL to integrate the views could be observed by comparing the uniform kernel and \tool's performances. The uniform kernel which assigns equal weights to all the base kernels performs on par with \tool{} obtaining best F-measure in {\tt CE1}. However, in {\tt CE2}, which involves larger datasets with more families of malware, the former fails to outperform the latter. Leveraging on MKL, our approach identifies the most appropriate linear combination of the base kernels and obtains better precision, recall and F-measure. This justifies the need for having a non-uniform combination of the base kernels. 
	
	\item Comparing against \soa{} approaches, we observe that \tool{} achieves performance on par with the best performing technique i.e., \textsc{Drebin}, while outperforming CSBD and \textsc{Adagio} in {\tt CE1}. It is important to note that \textsc{Drebin}'s features are perfectly engineered to offer excellent performance on the DR dataset. Similar observations have been reported through other large-scale studies such as \cite{acsac}. Hence achieving on-par performance convincingly reveals the detection potentials of \tool. Notably, all the approaches exhibit lesser F-measure in {\tt CE2} than in {\tt CE1}, revealing that the former setting is more challenging. Interestingly, \tool{} achieves best results in {\tt CE2} outperforming all the \soa{} approaches and the uniform kernel. 
	
\end{itemize}

Having sufficiently established the detection capabilities of base kernels and the need for MKL, we exclude evaluations and discussions on individual base kernels in forthcoming sub-RQ 1.2 and 1.3.

\subsubsection {RQ1.2 History-aware training and evaluation}
\label{sss:rq1.3}
We now report the results on experiment {\tt CE3}, where we enforced the constraint: \textit{the training set is historically anterior to the test-set} in Table \ref{tab:ce3}. 
\begin{table}[t]
	\centering
	\scriptsize
	\caption{{\tt CE3} Results: Precision, Recall and F-measure}
	\label{tab:ce3}
	\begin{tabular}{|l|C{1cm}|C{1cm}|C{1cm}|}
		\hline 
		Technique & P      & R      & F1     \\ \hline  \hline
		MKLDroid & \textbf{0.99} &  \textbf{0.55} & \textbf{0.71} \\ 
		Drebin \cite{Drebin} & \textbf{0.99} & 0.36 & 0.52       \\ 
		Adagio \cite{Adagio} & 0.98 & 0.41 & 0.58       \\ 
		CSBD \cite{CSBD} & 0.98 & 0.43 & 0.60        \\ \hline
	\end{tabular}
\end{table}
Before discussing the results of individual techniques, we present an important inference from Table \ref{tab:ce3}: the performances of all the state-of-the-art methods (and \tool) are significantly worse when the test-set is posterior to the training set. This is in-line with the observations reported in \cite{HM} and \cite{mama}. The reason for this drop in performance and relevant observations are reported below. 

Unlike experiments {\tt CE1} and {\tt CE2}, in {\tt CE3}, the test-set apps are almost one year historically posterior to the training set ones. Some of these malware could belong to a new family that emerged after the training set was compiled or could be a more sophisticated variant of a family that surfaced in the training set. Hence, all the knowledge acquired and used by the classifier would not be sufficient or relevant. 
	
In this aspect, {\tt CE3} models the real-world detection settings more closely. The practical potentials of a technique will be revealed in this experiment. Only techniques that capture most aspects of an app that distinguishes malicious behavior from benign ones could perform well. Techniques that overfit the training set will perform poorly particularly in {\tt CE3}.

From Table \ref{tab:ce3}, it is evident that \tool{} outperforms all the \soa{} techniques, significantly in {\tt CE3}. The margin of improvement is 11\%  which is  much higher compared to that of {\tt CE1} and {\tt CE2}. Furthermore, the following observations are made from Table \ref{tab:ce3}:

\begin{itemize} [leftmargin=*]
	\setlength\itemsep{0em}
	\item %\textit{Precision.} 
	Clearly the precision of all the techniques under comparison are very high and almost same (in the range [0.98,0.99]). This reveals that all the techniques could very well detect test-set malware that are similar to training set malware (e.g., similar variants of families exposed during training). 
	
	\item %\textit{Recall.} 
	However, the real difference between \tool{} and other techniques lies in the recall value. The other techniques have much poorer recall (well below 0.50). This reveals that even though they could detect known malware, they do not generalize well to find newer variants or families. \tool's recall value suggests that, it detects newer variants/families better. Nevertheless, its recall is also not very high, indicating high number of false negatives. This is because, though \tool{} can generalize well, it cannot automatically \textit{adapt} to the evolution in malware samples. More specifically,, \tool{} is a batch-learning based approach, where we use a batch of samples to train the model. Subsequent to training, the model is used only to predict the labels of the samples that stream-in. It cannot automatically update its learning unless it is retrained with a fresher or more recent training set. As noted in \cite{droidol} and \cite{prescience}, this is an inherent limitation of batch-learning based solutions. We discuss our measures to address this adaptiveness issue, later in \S \ref{sec:lim}. %{\color {red} Approaches like Prescience \cite{prescience} and DroidOL \cite{droidol} handle this problem by either periodically retraining the models or using online learning. We intend to investigate using online MKL \cite{omkl} in our framework as a future work}.
	
	\item Particularly, the recall and F-measure values of \textsc{Drebin}, which performed well in {\tt CE1} and {\tt CE2} are very poor in {\tt CE3}. This indicates that the features captured by \textsc{Drebin}, are too much dependent on the training set and it could reliably detect a new test-set malware only if it is very similar to training set malware. In other words, \textsc{Drebin} suffers from \textit{overfitting}. This observation is reinforced by the fact that some of \textsc{Drebin}'s features such as URLs, names of components, etc. do not adapt to unseen test-set apps well. For instance, the component names in test-set apps may be vastly different from training set ones and learning these becomes useless during testing. This introduces noisy features, leading to \textit{overfitting} and poor detection rates.
	
	\item Unlike \textsc{Drebin}, both \textsc{Adagio} and CSBD use features that are not very much training set-specific. This helps them generalize well. Furthermore, CSBD uses feature selection to prune irrelevant features and this helps improving its recall and F-measure.  However, these approaches perform more like signature-based detection approaches (as they learn instruction and CFG signatures) and this leads to an overall poor performance in {\tt CE3} where there is significant drift in the test-set. \tool{} outperforms these approaches by 11\% or more F-measure.
	
\end{itemize}

\subsubsection{RQ1.3 Accuracy on the wild dataset}
\label{sss:rq1.4}
We now report the detection results obtained on recent wild apps. In this experiment, {\tt WEx}, the models are trained using all the  malware from benchmark datasets and benign apps from Google Play and tested on apps collected from third-party markets. The precision, recall and F-measure values of all the approaches under comparison are presented in Table \ref{tab:itw}.

Evidently, in this experiment, \tool{} outperforms the \soa{} approaches significantly (i.e., by 8\% or more F-measure), similar to {\tt CE3}.
The following observations are made from Table \ref{tab:itw}.
\begin{itemize} [leftmargin=*]
	\setlength\itemsep{0em}
	
	\item At the outset, conspicuously, for all the approaches under study, performing detection on the wild dataset is more challenging than the benchmark ones. All the four methods produce F-measure in the range [0.60,0.72] in {\tt WEx}. This observation holds despite the fact that all models are trained with the largest amount of data in {\tt WEx}, out of all the four experiments. This reveals that the homogeneity observed in the attack vectors of benchmark malware is not observed in the same magnitude among malware apps in the wild. 
	%In fact, \cite{mist} reports that more than 85\% of the malware in DR dataset \cite{Drebin} perform privacy leaks related attacks. 
	%This observation of ours is in line with those made by recent works such as \cite{mist}. 
	
	\item %\textbf{Precision.} 
	Even though the F-measures of these approaches in {\tt WEx} are similar to those of {\tt CE3}, a major difference lies in the precision values. In {\tt CE3}, all the approaches identified test-set malware that exposed in the training set quite well, leading them to better precisions (i.e., in the range [0.98, 0.99]). However, in {\tt WEx}, their precision values are in the range [0.47, 0.60]. This on average is 45\% lesser than {\tt CE3}'s precision values. This clearly illustrates that the false positives are significant across all these approaches, as distinguishing the behaviors of benign and malware apps in the wild is more challenging.
	
	\item %\textbf{Recall.} 
	In {\tt CE3}, the test-set malware are entirely posterior to the training set ones, leading to a larger scope for malware evolution.  Consequently, all the four approaches obtained poor recall values (i.e., in the range [0.43,0.55]). However, in {\tt WEx}, the period of collection of test samples, overlaps with certain months of the training set compilation period. This enables the detection models to operate under a less vigorous population drift setting. This directly reflects in their lesser false negative and consequently, better recall values. {\color {black} However, we note these precision and recall values are not as high as in {\tt CE1} and {\tt CE2}, reinforcing the need for detecting and adapting to malware population drift observed in the wild.} 
	
	\item %\textbf{Vs. other approaches.} 
	Out of the \soa{} approaches, \textsc{Drebin} and CSBD perform reasonably better than \textsc{Adagio}. \tool{} outperforms these methods by 8\% or more F-measure. 
	
\end{itemize}

\begin{table}[t]
	\centering
	\scriptsize
	\caption{ {\tt WEx} Results: Precision, Recall and F-measure}
	\label{tab:itw}
	\begin{tabular}{|l|C{1cm}|C{1cm}|C{1cm}|}
		\hline 
		Technique & P      & R      & F1     \\ \hline \hline
		\tool{}                   &  \textbf{0.60}  &   \textbf{0.91} &   \textbf{0.72} \\ 
		Drebin \cite{Drebin}       &  0.53  &   0.78 &   0.63 \\ 
		Adagio \cite{Adagio}       & 0.47       &    0.87    &    0.61    \\ 
		CSBD \cite{CSBD}      &  0.51  &   0.87 &   0.64 \\ \hline
	\end{tabular}
\end{table}

\subsubsection {RQ1.4 Effectiveness of individual views}
\label{sss:rq1.2}
\begin{table}[t]
	\centering
	\scriptsize
	\captionsetup{justification=centering}
	\caption{Weights of \tool's base kernels - learnt using SMO-MKL \cite{SMO-MKL}}
	\label{tab:kernel-wts}
	\begin{tabular}{|c|c|c|}
		\hline
		\multirow{2}{*}{\textbf{Kernel}} & \textbf{{\tt CE1}}    & \textbf{{\tt CE2}}    \\ 
		& Weight (norm = L2, C = 100) & Weight  (norm = L2, C = 1) \\ \hline \hline
		{\tt API} kernel                         &   \textbf{119.88 ($\pm$ 1.16)}    &   \textbf{29.66 ($\pm$ 0.18)}   \\ 
		{\tt Permission} kernel                        &    46.41 ($\pm$ 0.73)    &   13.37 ($\pm$ 0.12)   \\ 
		{\tt Src-sink kernel}                       &   26.41 ($\pm$ 0.80) 	    &   12.45 ($\pm$ 0.15)   \\ 
		{\tt Instruction} kernel   &   95.34 ($\pm $1.09)
		&   16.27 ($\pm $0.81)   \\ 
		{\tt Signature} kernel     &   95.77 ($\pm$ 1.31)
		&   14.26 ($\pm$ 0.71)   \\ \hline
	\end{tabular}
\end{table}

Having sufficiently established that capturing context-aware multi-view features facilitates \tool{} to achieve superior accuracies, we now intend to gain insights into the contributions of individual views to its performance.
In this sub-RQ, we perform a sensitivity analysis on \tool's base kernels both quantitatively and qualitatively. In particular, we study whether there is any correlation between the weights that MKL offers to individual base kernels and their detection rate, rank the views based on these weights and also suggest which views to use when resources and time are at a premium.

\textbf{Quantitative analysis.} As explained in \S \ref{ss:mkl}, the SMO-MKL algorithm used in \tool{} assigns weights to individual base kernels yielding an appropriate linear combination of the base kernels. These weights of the base kernels in {\tt CE1} and {\tt CE2} are reported in Table \ref{tab:kernel-wts}. The following observations are made from the table:
\begin{itemize} [leftmargin=*]
	\setlength\itemsep{0em}
\item A straight-forward observation is that, in our framework, the weight of a base kernel signifies its malware detection potential. Higher the weight, better is the kernel’s detection potential. For instance, the \texttt{API} and \texttt{Src-sink} kernels get the most and least significant weights in both the experiments. This is in line with the fact that these kernels offered the best and the worst standalone accuracies, respectively. However, we note that there need not be direct correlation between the base kernel weights and their accuracies when non-linear kernel combinations or hyper-kernels are used. Our base kernels' weight assignments are particularly intuitive as we perform linear MKL in our framework.

\item Also, the weights of {\tt Instruction} and {\tt Signature} kernels are very similar. This stems from the fact that these kernels capture similar structural information at similar level of granularity, exhibiting identical detection capabilities. Interestingly, {\tt Signature} kernel obtains more weight than {\tt Instruction} kernel in {\tt CE1}, whereas the weight significances are other way around in {\tt CE2}. We believe this due to the fact that the dataset used in {\tt CE1} (i.e., DR) is comparatively more homogeneous than in {\tt CE2} (i.e., VS). Homogeneous malware tend to exhibit similar CFG signature thereby helping the {\tt Signature} kernel to get a trifle better detection rates. 

\item {\tt Permission} kernel's weights rank just below the {\tt Signature} and {\tt Instruction} kernels, going hand-in-hand with its prediction quality.

%\item \color{red}{Finally, we observe that the standard deviations of kernel weights in {\tt CE2} across 5 runs is larger that those of {\tt CE1}. This again reinforces that across different runs, all the kernels exhibit closer or more repeatable performances in the latter setting compared to the former one.} 

\end{itemize}

\textbf{Qualitative analysis.} Due to space limitations, the results of qualitative analysis (and visualization of kernel matrices) that reinforce those of quantitative analysis are presented in Appendix \ref{app:qual}.

\textbf{Ranking the individual views.} With the explanations mentioned above, it is straight-forward to rank the kernels. Considering that the base kernel weights quantifies their relative importances in the MKL setting, the following rankings are obtained: rank 1: {\tt API} kernel, rank 2: shared by {\tt Signature} and {\tt Instruction} kernels, rank 4: {\tt Permission} kernel and rank 5: {\tt Src-sink} kernel. However, the scalability of these kernels differ significantly. Hence we defer discussing which subset of kernels to use, when time/resources are scarce, to \S \ref{ss:eff}, where we report the base kernel efficiencies.\\

{%group to keep \parindent change local
	\parindent-\fboxsep     %revert indentation due to \fbox frame space
	\indent%    
	\fbox{%
		\parbox{\linewidth}{%
%			\parindent\defaultparindent%
			\indent \textit{Summarizing the inferences from RQ1, we conclude that individual base kernels of \tool{} exhibit sub-par malware detection potentials and combining them appropriately through MKL allows them to complement each other to achieve superior accuracies. This context-aware multi-view learning makes \tool{} powerful enough to detect more sophisticated and newer variants of malware, leading it to significantly outperform \soa{}  methods in challenging real-world experimental settings.}
		}
	}
}%end parindent group

\subsection {RQ2: Efficiency}
\label {ss:eff}

We now present the results for efficiency of our base kernels, \tool{} and \soa{} approaches in terms of average training and testing durations in experiments {\tt CE1} and {\tt CE2} (across 5 runs) in Table \ref{tab:eff}. 
The training and testing time depends on factors such as sample size, number of features and type of the kernel/learner being used. Hence, these values are also reported. It is noted that the trend in efficiency values remain same in other experiments (i.e., {\tt CE3} and {\tt WEx}) as well. Owing to space constraints, they are omitted.

The following are noted as well: (i) for the base kernels that involve feature selection ({\tt Instruction}  \& {\tt Signature}) the reported training duration includes time taken for feature selection, corresponding dimensionality reduction and training the models, and (ii) in the case of \tool{}, testing duration includes time taken for switching from dual to primal formulation and then predicting the labels (see \S \ref{ss:lm}).
%Also, as \tool{} and \textsc{Drebin} use BoF model, their number of features will vary for every instance of training based on the samples in the training set. Hence, we report the average number of features across 5 runs. 
Expectedly, {\tt CE2}'s training and testing durations are longer than {\tt CE1} as it involves larger datasets. From Table \ref{tab:eff} the following inferences are drawn:
\begin{itemize} [leftmargin=*]
	\setlength\itemsep{0em}
	\item Since PRG vectors from individual views (e.g., CADG vectors for {\tt API} kernel, and so on) are used in conjunction with linear SVMs, their base kernel training and testing durations is very less. Since, both uniform kernel and MKL perform classification through linear combinations of base kernels, they require slightly more training and testing durations.
	
	\item \textbf{Comparing individual views.} Out of the base kernels, {\tt Src-sink} and {\tt Permission} use least number of features and hence emerge as the fastest ones in terms of both training and testing durations. On the other hand, {\tt API} kernel uses very large number of features and hence emerges as the slowest base kernel. As mentioned earlier in \ref{ss:fe}, in the case of {\tt Instruction} and {\tt Signature} kernels, we obtain extremely large number of features (i.e., more than 500,000) and hence feature selection is used choose and retain only 5,000 most informative features. This results in significantly larger training and testing durations. In summary, in terms of efficiency, \tool's base kernels could be ranked as follows: rank 1: {\tt Src-sink}, rank 2: {\tt Permission}, rank 3: is shared by both {\tt Instruction} and {\tt Signature} and rank 5: {\tt API}. Evidently, when we take into account accuracy results from tables \ref{tab:ce1} and \ref{tab:ce2} as well, the two most efficient views are not sufficiently accurate. Hence, when there are severe resource/time constraints, we recommend using only the {\tt Instruction} and/or {\tt Signature} kernels which achieve high accuracy with reasonable efficiency. Following the above-mentioned strategy, the accuracy-efficiency trade-off among the individual views of \tool{} could be learnt from tables \ref{tab:ce1}, \ref{tab:ce2} and \ref{tab:eff}. For particular time or resource constraints, one may decide on which views to use, based on this trade-off.
	
	\item \textbf{Vs. \textsc{Drebin}.} \textsc{Drebin} extracts light-weight features (i.e, no PRG based features are extracted) and uses linear SVM for classification. This helps \textsc{Drebin} to achieve very high efficiency. Comparing \textsc{Drebin}'s efficiency against those of  \tool's base kernels, we could see that all of them are much faster than \textsc{Drebin}. This is mainly because these base kernels use lesser number of features than \textsc{Drebin}. Uniform kernel's efficiency is comparable to that of \textsc{Drebin}. However, none of the base kernels could outperform \textsc{Drebin} in terms of accuracy. Meaning, base kernels achieve this efficiency at the cost of accuracy. Also, when the base kernels are integrated, \tool{} becomes almost 5 to 6 times slower in terms of training duration and 42 to 44 times slower in terms of testing duration than \textsc{Drebin}. This is mainly because of the time \tool{} spends to learn the base kernel weights using SMO-MKL and switching to primal formulation, subsequently.
	
	\item \textbf{Vs. CSBD.} CSBD uses RF classifier with 100 estimators and 5000 features (selected using Information Gain values). Since RFs are quasi-linear models they require significantly more training and testing durations than approaches that use linear models (i.e., \textsc{Drebin} and \tool).  In particular, \tool{} is more than 2 times faster in terms of training and more than 17 times faster in terms of testing durations than CSBD.
	
	\item \textbf{Vs. \textsc{Adagio}.} \textsc{Adagio} uses a kernel SVM, a computationally heavy learner, as it aims to learn non linear decision boundaries.  Furthermore, it uses  HI kernel with large number of  features in the primal formulation, as it aims to build an  interpretable model. These factors render the approach very much inefficient with practically intractable training and testing durations. In particular, \tool{} is more than 595 times faster in terms of training and more than 10,924 times faster testing durations than \textsc{Adagio}. 
\end{itemize}

\begin{table*}[t]
	\centering
	\scriptsize
	\setlength\tabcolsep{2pt}
	\caption{{\tt CE1} and {\tt CE2} Efficiency: Average training and testing durations}
	\label{tab:eff}
	\begin{tabular}{|c|c|ccc|ccc|}
		\hline
		\multirow{2}{*}{\textbf{Technique/Kernel}} & \multirow{2}{*}{\textbf{\begin{tabular}[c]{@{}c@{}}Kernel\\ Type\end{tabular}}} & \multicolumn{3}{c|}{\textbf{{\tt CE1}}} & \multicolumn{3}{c|}{\textbf{{\tt CE2}}} \\ \cline{3-8} 
		&  & \textbf{\#  features} & \textbf{\begin{tabular}[c]{@{}c@{}c@{}}Feat. Select.\&\\Tr. duration\\ (in sec.)\end{tabular}} & \textbf{\begin{tabular}[c]{@{}c@{}c@{}}Testing \\duration\\ (in sec.)\end{tabular}} & \textbf{\# features} & \textbf{\begin{tabular}[c]{@{}c@{}c@{}}Feat. Select.\&\\Tr. duration \\ (in sec.)\end{tabular}} & \textbf{\begin{tabular}[c]{@{}c@{}c@{}}Testing \\duration\\ (in sec.)\end{tabular}} \\ \hline \hline
		{\tt API} & Linear & 102633 & 0.56 & 0.0013 & 115380 & 0.63 & 0.0014 \\ 
		{\tt Permission} & Linear & 253 & 0.02 & 0.0001 & 367 & 0.02 & 0.0001 \\ 
		{\tt Src-sink} & Linear & 129 & 0.01 & 0.0001 & 147 & 0.02 & 0.0001 \\ 
		{\tt Instruction} & Linear & 5000 & 0.36 & 0.0003 & 5000 & 0.38 & 0.0004 \\ 
		{\tt Signature} & Linear & 5000 & 0.33 & 0.0002 & 5000 & 0.37 & 0.0004 \\ \hline
		Uniform kernel & \begin{tabular}[c]{@{}c@{}}Linear combination \\ of base kernels\end{tabular} & 113015 & 0.73 & 0.0016 & 125894 & 0.81 & 0.0016 \\ \hline
		\tool\ & \begin{tabular}[c]{@{}c@{}}Linear combination \\ of base kernels\end{tabular} & 113015 & 4.67 & 0.0725 & 125894 & 5.30 & 0.0870 \\ \hline
		\textsc{Drebin} \cite{Drebin} & Linear & 170185 & 0.75 & 0.0017 & 390431 & 0.98 & 0.0020 \\ \hline
		CSBD \cite{CSBD} & N/A & 5000 & 13.62 & 1.24 & 5000 & 46.89 & 1.34 \\ \hline
		\textsc{Adagio} \cite{Adagio} & HI kernel & 32768 & 2783 & 792 & 32768 & 5188 & 961 \\ \hline
	\end{tabular}
\end{table*}

\leavevmode 
\newline
{%group to keep \parindent change local
	\parindent-\fboxsep     %revert indentation due to \fbox frame space
	\indent%    
	\fbox{%
		\parbox{\linewidth}{%
			%			\parindent\defaultparindent%
			\indent \textit{Summarizing RQ2 evaluations, we conclude, though \tool{} uses five base kernels, it sticks to a linear combination of the kernels and hence requires modest training and testing durations. It is more efficient than two of the \soa{} approaches. Furthermore, the trade-off between base kernels' accuracy and efficiency is determined, which would help in using a select few of them when resources and time constraints are severe.}
		}
	}
}%end parindent group

%ADG kernel and the SSCFP kernels obtain the highest and the lowest weights, consistently across both the experiments. This indicates that, out of the individual base kernels, ADG kernel has the most reliable prediction capability. This inference is reinforced by the findings presented in Table \ref{tab:ce12}, where the ADG kernel offered the best the accuracy among the base kernels. SSCFP kernel gets the least weight as its individual accuracy is relatively poor.

\subsection {RQ3: Locating Malice Code}
\label {ss:lme}
In this RQ, we intend to investigate whether \tool{} is capable of reliably locating malice code in a given sample both qualitatively and quantitatively. As stated earlier, all the existing approaches\footnote{Though \textsc{Adagio}, in principle could identify malice methods from CGs, the implementation provided at \cite{adagiocode} does not include this.} (incl. \textsc{Drebin} and CSBD) are not capable of doing such localization. Since, fine-grained malicious code localization is \tool’s unique feature, we could not compare this with any existing technique. Hence, we illustrate how \tool{} achieves it in this subsection and discuss how it supports human analysts to visualize PRGs from different perspectives facilitating precise understanding of malice behaviors

\subsubsection {Qualitative Evaluation}
\begin{figure*}
	\includegraphics[height=20cm,width=15cm]{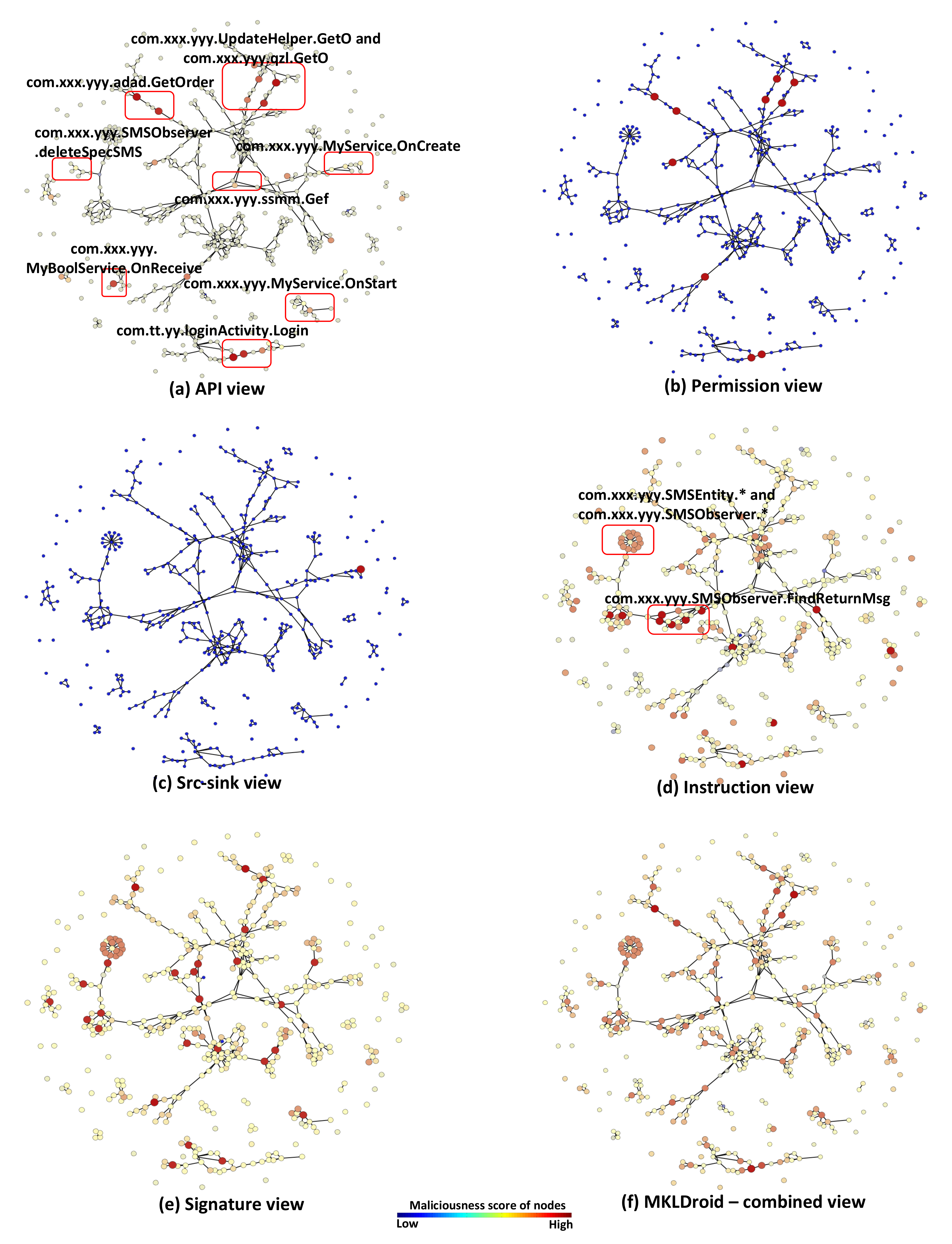}
	\caption{Depictions of CICFG of the \textit{ADRD}  sample from five different views.  (a) {\tt API} view: the CICFG nodes are assigned m-scores according to their CADG features i.e., {\tt API} m-score (b) {\tt Permission} view (c) {\tt Src-sink} view (d) {\tt Instruction} view (e) {\tt Signature} view (f) MKL view: for every CICFG node, the m-scores assigned are MKL based linear combinations of their individual views' m-scores. \label{fig:adrd}}
\end{figure*}

To perform qualitative evaluation we trained the model as in experiment {\tt CE1} and use it to locate malice code in the test-set apps. Once a test-set app is  predicted to be malicious, \tool{} assigns m-scores to each CICFG nodes (i.e., basic block) as described in \S \ref{ss:lm}. Hence, for the qualitative analysis part, we manually investigate the code in basic blocks with high m-scores to check whether they indeed are a part of the malware's attack vector. 

The above mentioned analysis yielded interesting and useful results for a substantial majority of the test-set apps.
We choose to explain the results of two popular malware families, \textit{ADRD} and \textit{Geinimi}. \textit{ADRD} is a family of malware that was wide-spread during the earlier versions of Android, had simple attack vectors to perform privacy leaks and were not repackaged malware. Meaning, almost all the code in the \textit{ADRD} samples are mal-intended. On the other hand, \textit{Genimi} performs more sensitive privacy leak attacks with more sophistication. Also, \textit{Geinimi} samples in the DR dataset are repackaged versions of popular benign apps. Meaning, only a minority of the code in the \textit{Geinimi} samples are mal-intended. 

\textbf{Case Study 1: \textit{ADRD} .}
The \textit{ADRD}  sample\footnote{MD5: \scriptsize{ 1944d8ee5bdda3a1bd06555fdb10d3267ab0cc4511d1e40611baf3ce1b81e5e8}} used in this study attempts to perform the following malice actions in the background after the phone is booted: accessing users' personal sensitive information (e.g., IMEI, IMSI and network information) and sending them to remote servers, sending and deleting SMS messages, downloading unsolicited apps, and issuing HTTP search requests.

The five different views derived from the app's CICFG are presented in figures \ref{fig:adrd} (a) to (e). For the {\tt API} kernel's view (in Fig. \ref{fig:adrd} (a)), the CICFG nodes are awarded m-scores based only the CADG contextual subgraph features that emerge from them. Similar m-score assignments are done to CICFG nodes in the remaining views in sub-figures (b) to (e). Finally, MKL based m-scores (which is a weighted sum of individual views' m-scores) are assigned to the CICFG nodes in Fig. \ref{fig:adrd} (f). The nodes are scaled in size and colored according to their m-score. Larger/warmer (i.e., reddish) nodes denote nodes with high m-scores which are potentially malicious and smaller/cooler (i.e., blueish) denote nodes that supposedly do not involve in attack related activities.

The following observations are made from Fig. \ref{fig:adrd} (a):
\begin{itemize} [leftmargin=*]
	\setlength\itemsep{0em}
\item From {\tt API} view, we could see that several basic blocks from classes such as \textit{com.xxx.yyy.MyService}, \textit{com.xxx.yyy.MyBoolService} and \textit{com.xxx.yyy.adad.GetOrder} get high {
\tt API} m-scores. We inspected each of these methods manually and realized that these methods indeed perform malice activities. The precise dissection is presented below:

\item \textbf{Trigger.} Similar to many popular malware families this \textit{ADRD}  sample uses broadcast notifications and alarm manager APIs to trigger its malicious operations. More specifically, method \textit{com.xxx.yyy.MyBoolService.OnReceive} uses {\tt Intent.getBroadcast} API to listen to a specific broadcast message notifying the {\tt BOOT\_COMPLETED} event. Once it receives the  message, it sets an alarm that is fired periodically using the   {\tt AlarmManager.set} API. These alarms would start a background service named \textit{MyService}. 

\item \textbf{Reading private data.} \textit{MyService} begins its lifecycle execution through invoking the \textit{com.xxx.yyy.MyService.OnCreate} and \textit{com.xxx.yyy.MyService.OnStart} methods.
%\footnote{If the service is triggered for the first time, it will call \textit{onCreate} and \textit{onStart}; otherwise, it will only call \textit{onStart}}. 
In the \textit{OnCreate} method, the IMEI and IMSI numbers are collected by invoking {\tt getDeviceId} and {\tt getSubsriberId} APIs. Also, it registers an object handler to access the SMS database ({\tt content://sms/}). The \textit{OnStart} method collects some more private information such as network information (e.g., type of the network |'wifi' or 'UNIWAP') by invoking {\tt getActiveNetworkInfo} and {\tt getTypeName} APIs.

\item \textbf{Leaking private data.} Once all this information is collected, the sample encrypts and leaks them over the internet. This is done in methods \textit{com.xxx.yyy.adad.GetOrder}, \textit{com.xxx.yyy.UpdateHelper.GetO} and \textit{com.xxx.yyy.qzl.GetO}. The methods invoke sensitive APIs such as {\tt DefaultHttpClient.init, java.io.FileOutputStream.write} and {\tt DefaultHttpClient.execute}. Alternatively, the same information is exfiltrated through SMS using the APIs {\tt SmsManager.getDefault} and {\tt sendTextMessage} in method \textit{com.xxx.yyy.ssmm.Gef}.

\item \textbf{Reading and deleting SMS.} Besides this, the service uses method \textit{xxx.yyy.SMSObserver.deleteSpecSMS} to monitor changes to the SMS database by calling {\tt ContentObserver.onChange}  API and deleting particular messages using {\tt ContentResolver.delete} API.

\item \textbf{Downloading unsolicited apks.} Moreover, the sample attempts to download a new unsolicited apk named '\textit{myupdate.apk}' and install the same on the device in method \textit{com.tt.yy.loginActivity.Login}. 

\end{itemize}

All these operations involved invoking sensitive APIs and hence are adequately captured in the {\tt API} view. We now turn our attention to other remaining views. 
The following inferences from figures \ref{fig:adrd} (b) to (e):

\begin{itemize} [leftmargin=*]
	\setlength\itemsep{0em}

\item The {\tt Permission} view (fig. \ref{fig:adrd} (b)) depicts the CICFG nodes scaled according to CPDG features based m-scores. As mentioned earlier, this \textit{ADRD}  sample's attack is simple and it turns out that only four permissions are required to carry out this, namely, {\tt READ\_PHONE\_STATE} (for reading IMEI, IMSI etc.), {\tt ACCESS\_NETWORK\_STATE} (for reading the network type), {\tt INTERNET}(for leaking information through the network and communicating to C\&C server) and {\tt SEND\_SMS} (for leaking information through SMS). 
Evidently, much less number of nodes correspond to code that uses these permissions and only these nodes get significant {\tt Permission} m-scores. This is why we could see a lot of cooler/smaller nodes in this view. Out of the permissions, {\tt INTERNET} is leveraged by almost all the \textit{ADRD}  variants for leaking information and are more popular than leaks via SMS. Consequently, the CPDG features related to former permission get higher m-scores than the latter ones. 
The features related to other two permissions are assigned lesser m-scores as well.
Interestingly, the APIs used to read and delete SMS are not permission protected and hence they are assigned insignificant m-scores. Overall, in this view, we have only a few suspicious nodes and they are from the following methods: \textit{com.xxx.yyy.UpdateHelper.GetO}, \textit{com.xxx.yyy.adad.GetOrder}, \textit{com.xxx.yyy.qzl.GetO} and \textit{com.tt.yy.loginActivity.Login}.

\item The {\tt Src-sink} view (fig. \ref{fig:adrd} (c)) depicts the CICFG nodes scaled according to {\tt Src-sink} m-scores. There exists only one sensitive control flow path in this CICFG and it originates from \textit{com.xxx.yyy.MyService.OnCreate} method. This path connects the source {\tt UNIQUE\_IDENTIFIER} to sinks {\tt FILE}\footnote{In this context, the leaks through internet is considered akin to writing into a file and hence we see a {\tt FILE} sink instead of a {\tt NETWORK} sink.} and {\tt SMS}. We note that this sample leverages ICCs heavily. As our method does not capture ICCs and some paths that exist between other sources such as content resolver and {\tt FILE}/{\tt SMS} might be missed by our approach. Overall, in this view, we have only one suspicious node.

\item Evidently, the {\tt Permission} and {\tt Src-sink} views over-abstract the semantics of the samples compared to the {\tt API} view. Moreover, in the case of this sample, the information captured by the two former views is a supplementary to that captured in the {\tt API} view. In other words, this example clearly reinforces the finding from RQ1 that the level of abstraction attained in the {\tt Permission} and {\tt Src-sink} views are too coarse-grained to be effective. %As a result, many nodes that perform malice operations get lesser scores and look smaller/cooler in these views, compared to CADG.

\item As in figures \ref{fig:adrd} (d) and (e), the CICFGs scaled according to the {\tt Instruction} and {\tt Signature} m-scores and  look very much different from the three aforementioned views. This is expected as those three views are closely related and capture similar semantics of the app. However, {\tt Instruction} and {\tt Signature} views, as discussed before, footprint the apps more like syntax-based detectors rather than semantics-based ones. For instance, the methods considered as significant in these two views are ones that belong to classes \textit{com.xxx.yyy.SMSEntity}, \textit{com.xxx.yyy.SMSObserver}. These two classes are present in the same composition in more than 27\% of the \textit{ADRD}  samples in the DR dataset. 
These methods predominantly contain \textit{utilities} code that is common across multiple variants of the \textit{ADRD}  family. 
Meaning, their CFG signatures and instruction sequences remain same across multiple samples and act as good features to characterize code that is unique in malware samples, but not necessarily malice. These two views precisely exploit these features to detect \textit{ADRD}  footprints. Figures \ref{fig:adrd} (d) and (e) visually illustrate that these two views capture information that is complementary to the three other views. 

\item Finally, we scale the nodes according to  \tool's m-scores, which is nothing but the linear combinations of their m-scores in each of the five views. 
%Each view gets m-scores according to its malware detection potential and the final combination of all of them is presented in Fig. \ref{fig:adrd} (e). 
Evidently, this multi-view representation retains useful information from all the views. For instance, this view considers both \textit{com.xxx.yyy.UpdateHelper.GetO} which is significant due to its semantic functionalities and \textit{com.xxx.yyy.SMSObserver.FindReturnMsg} which is significant due to its popularity across multiple \textit{ADRD}  variants as more or less equally malicious. 

\item Finally, we note that this sample is not piggybacked. Meaning, much of the code (i.e., CICFG nodes) take part in the attack performing potentially harmful operations. This is indeed reflected well when we visualize \tool's results as most nodes are larger and warmer looking (i.e., assigned high m-scores). %\tool{} achieves this by heavily borrowing from its three most useful views (i.e., CADG, {\tt Instruction} and {\tt Signature}). 

\end{itemize}

\begin{figure*}
	\includegraphics[height=20cm,width=15cm]{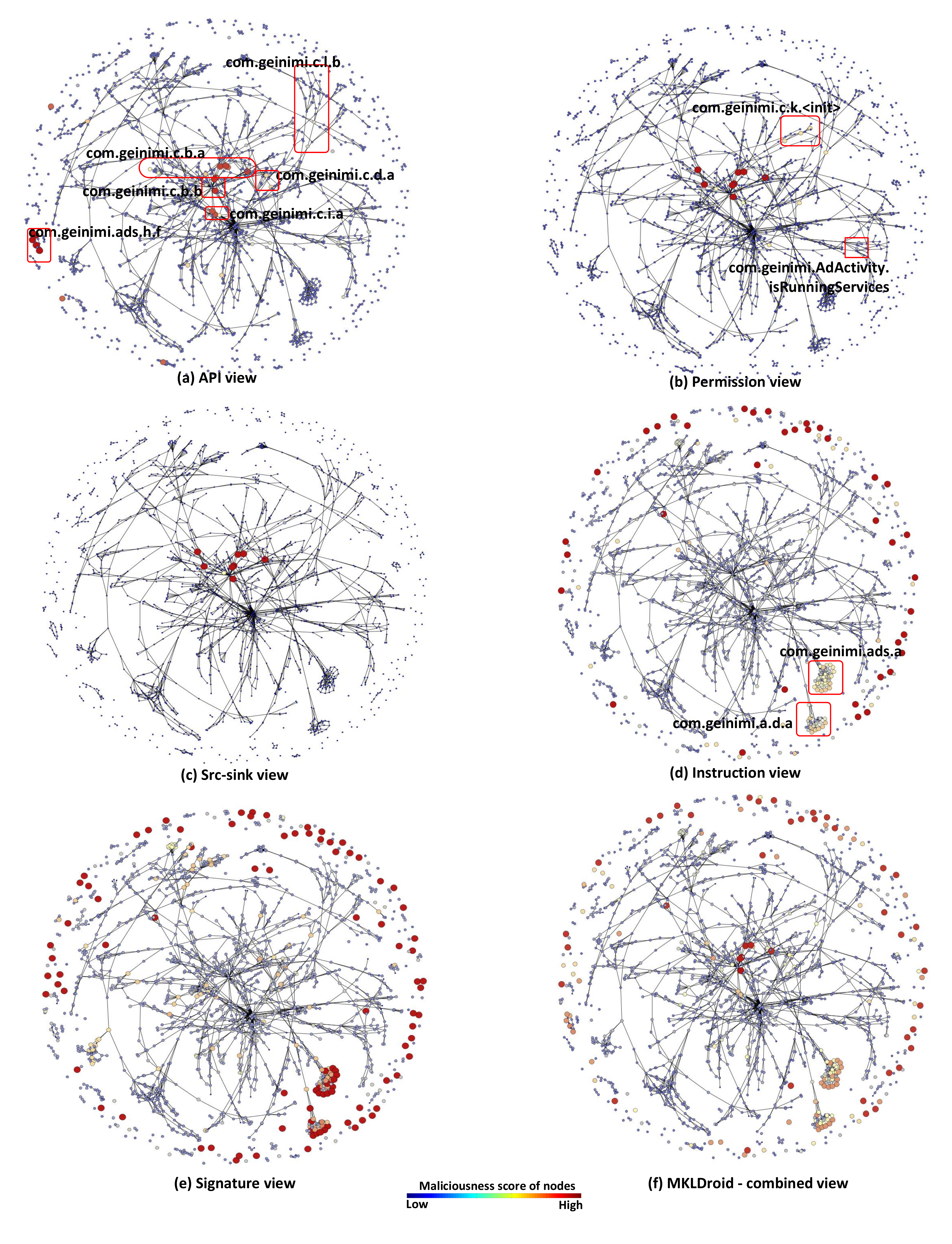}
	\caption{Depictions of CICFG of the \textit{Geinimi} sample from five different views. \label{fig:geinimi} (a) {\tt API} view: the ICFG nodes are assigned m-scores according to their CADG features i.e., {\tt API} m-score (b) {\tt Permission} view (c) {\tt Src-sink} view (d) {\tt Instruction} view (e) {\tt Signature} view (f) MKL view:for every CICFG node, the m-scores assigned are MKL based linear combinations of their individual views' m-scores.}
\end{figure*}

\textbf{Case Study 2: \textit{Geinimi}.}
The \textit{Geinimi} sample\footnote{MD5: \scriptsize{7bbd566f2f3abb78b3ffcc23ba4ad84e06a00f758d245c660c61b21814a850a5}} used in this case study is the real-world version of the working example presented in \S \ref{sec:bgm}. Unlike the \textit{ADRD}  sample, this is a repackaged malware. This sample's malice code is piggybacked on a popular benign game app. Also, the malicious functionalities in this sample are more sophisticated than \textit{ADRD} . \textit{Geinimi}'s malice code is triggered through a system-generated broadcast message. On receiving it, \textit{Geinimi} starts a service in the background. This service read volumes of personal data such as the users' location, contacts, emails and leaks them through internet and SMS messages.

The five different views derived from the app's CICFG are presented in figures \ref{fig:adrd} (a) to (e). At the outset, it is evident that unlike \textit{ADRD} , across all the views, only a few nodes actually potentially malicious and have received high m-scores. This is because in the case of this piggybacked Genimi app, only a few nodes involve in mal-intended operations (i.e., rider code) and the majority of the code is benign (i.e., host app's code). 

The following observations are made from Fig. \ref{fig:geinimi} (a):
\begin{itemize} [leftmargin=*]
	\setlength\itemsep{0em}

\item \textbf{Trigger.} On receiving the {\tt BOOT\_COMPLETED} event notification, a service named \textit{com.geinimi.AdService} is started. This service continues to collect a variety of private information and leaks the same as described below.

\item \textbf{Collecting location information.} \textit{Geinimi} collects user's geographic location in method \textit{geinimi.c.d.a} using APIs {\tt getLastKnownLocation}, {\tt getLatitude} and {\tt getLongitude}. 

\item \textbf{Collecting device identifiers.} The method \textit{geinimi.c.k.init} collects 13 different types of private information including phone number, IMEI, IMSI, mobile service operator's name, etc. using several APIs such as {\tt getLine1Number, getSimSerialNumber, getSimOperatorName}, etc.

\item \textbf{Collecting contacts.} This sample attempts to read users' contacts stored in the content providers through invoking {\tt getContentResolver} and {\tt ContentResolver.query} APIs. These operations are performed in methods \textit{geinimi.c.b.a} and \textit{geinimi.c.b.b}. The former method just reads the contact's display name, last contacted time and phone number. The latter method collects  email addresses along with the above-mentioned information and bundles the same with the device's unique identifier making them ready to be leaked.

\item \textbf{Collecting emails related information.} The emails stored in the content resolver are read using several methods in the class \textit{geinimi.ads.h} and method \textit{f} in this class bundles the emails' \textit{to} address, \textit{cc/bcc} address list, \textit{subject} and content into an intent message (using ICC related APIs such as {\tt Intent.putExtra}) and send them across to other methods for leaking them over internet.

\item \textbf{Leaking over internet and SMS.} Finally, this sample leaks the collected private information over the internet in method \textit{geinimi.c.l.b} using APIs such as {\tt java.net.HttpURLConnection.init, java.net.URL.init, URL.openconnection, java.io.DataOutputStream.write, flush, close} and {\tt HttpURLConnection.disconnect}. Alternatively, the same private contents are leaked through SMS in method \textit{geinimi.c.i.a}. This method uses APIs {\tt SMSManager.getDefault} and {\tt sendTextMessage}.
\end{itemize}

Similar to the case of \textit{ADRD}'s {\tt API} view, all these operations in \textit{Geinimi} involved invoking sensitive APIs and hence are adequately captured in this view. We now make the following observations from figures \ref{fig:geinimi} (b) to (f):
\begin {itemize}[leftmargin=*]
\setlength\itemsep{0em}
\item Similar to the {\tt Permission} view in the previous case study, only a minority of \textit{Geinimi}'s APIs/URIs are permission protected and consequently, receive high {\tt Permission} m-scores. Predominantly, the following permissions are used by this sample to carry out the privacy leaks: {\tt ACCESS\_COARSE\_LOCATION, ACCESS\_FINE\_LOCATION, READ\_CONTACTS, READ\_PHONE\_STATE, INTERNET} and {\tt SEND\_SMS}. All these permissions are leveraged by the methods discussed above. Interestingly, since the method \textit{geinimi.c.k.init} involves accessing a large number of device identifier related APIs across its different basic blocks, it happens to use {\tt READ\_PHONE\_STATE} permission repeatedly. This behavior has rewarded more significant weights to these basic blocks in the {\tt Permission} view than the CADG one. Also, another method, \textit{com.geinimi.AdActivity.isRunningServices}, that helps to obfuscate \textit{Geinimi}'s attacks through using APIs such as {\tt Class.forName} and {\tt getClassName}. These are not permission protected APIs, however, we mapped them to special permissions (see \S \ref{ss:sa}) and this has yielded the corresponding nodes, more significant m-scores in the {\tt Permission} view compared to other views.

\item Since \textit{Geinimi}'s privacy leaks involve more variety of information sources being read and leaked than \textit{ADRD} , we could see several nodes with large m-scores in the {\tt Src-sink} view in Fig. \ref{fig:geinimi} (c). Typically, these control flow paths originate from the following methods: \textit{com.geinimi.c.b.a, com.geinimi.c.b.b} and \textit{com.geinimi.c.d.a}. 

\item Similar to the observations made in \textit{ADRD} , both the {\tt Permission} and {\tt Src-sink} views over-abstract \textit{Geinimi}'s attacks. For instance, the sensitive operations in methods \textit{com.geinimi.ads.h.f} and \textit{com.geinimi.c.l.b} are not adequately captured in both these views. However, these two views could capture some critical information which is not reflected in {\tt API} views. For instance, they exclusively reveal the sensitive operations in methods such as \textit{com.geinimi.AdActivity.isRunningServices}.

\item Similar to the \textit{ADRD}  sample's case, the {\tt Instruction} and {\tt Signature} views footprint methods are statistically prominent across multiple variants of the \textit{Geinimi} family. Specifcially, two methods \textit{com.geinimi.ads.a} and \textit{com.geinimi.ads.b} which figure in more than 15\% of the \textit{Geinimi} samples in the same composition are leveraged to footprint them in both these views. 

\item Finally, all the aforementioned views are integrated using MKL in Fig. \ref{fig:geinimi} (f). Evidently, semantically significant nodes from methods such as \textit{com.geinimi.c.b.a} and utility code related nodes from methods such as \textit{com.geinimi.ads.a} receive high m-scores thus capturing the best of both worlds, helping effective detection. Overall, since this app is piggybacked with a large amount of benign code, one could see a large number of cooler/smaller nodes in \tool's multi-view CICFG. This is in sharp contrast from \textit{ADRD} 's multi-view CICFG in Fig. \ref{fig:adrd} (f).

\end {itemize}

\subsubsection{Quantitative Evaluation}
\label{sss:lme_quant}

\begin{table}[t]
	\centering
	\setlength\tabcolsep{5pt}
	\scriptsize
	\caption{{\tt MCLEx}: Quantitative Evaluation of malicious code localization on the MYST dataset}
	\label{tab:lme}
	\begin{tabular}{|c|c|c|c|c|}
		\hline
		 \begin{tabular}[c]{@{}l@{}}\textbf{Avg. \# of} \\ \textbf{malice classes} \end{tabular} & \begin{tabular}[c]{@{}l@{}}\textbf{FPR @ 10} \\ \textbf{(avg. $\pm$ std.)}\end{tabular} & \begin{tabular}[c]{@{}l@{}}\textbf{FNR @ 10} \\ \textbf{(avg. $\pm$ std.)}\end{tabular} &
		\begin{tabular}[c]{@{}l@{}}\textbf{P @ 10} \\ \textbf{(avg. $\pm$ std.)} \end{tabular} &
		\begin{tabular}[c]{@{}l@{}}\textbf{R @ 10} \\ \textbf{(avg. $\pm$ std.)} \end{tabular} \\ \hline
		2.48 & 17.00 ($\pm$ 1.60) & 5.67 ($\pm$ 20.16) & 14.34 ($\pm$ 7.99) & 94.33 ($\pm$ 20.16) \\ \hline
	\end{tabular}
\end{table}

We now present the results of quantitative evaluation of our experiments to locate malice code from MYST dataset. As mentioned earlier, in this experiment, the model is trained using 2,000 MYST malware and 2,000 benign apps. This model is then allowed to locate malice classes in each of the remaining 1,000 MYST apps. We consider classes with highest MKL based m-scores as malicious and compare them against the ground-truth on names of malice classes to compute the correctness and completeness of detection.

On average, we have 70.38 classes in MYST apps and according to ground truth, 2.48 of them are malicious. Given this statistics, we consider that malice code is indeed present in top 10 classes with highest m-scores awarded by \tool{}. Meaning for an analyst working on this dataset, \tool{} would reduce the search space for seeing malice code by 1/7th of the total code, on average. 
FPR, FNR, precision and recall values averaged across all the 1,000 appss are presented in Table \ref{tab:lme}. The following inferences are drawn from the table:
\begin{itemize} [leftmargin=*]
	\setlength\itemsep{0em}
\item The average FNR is just over 5\% and the recall values is more than 94\%. Meaning, while inspecting only 1/7th of the code, \tool{} facilitates the analysts to detect and inspect more than 94\% of all the classes that are involved in malice operations. This is immensely helpful for analysts as it narrows down their search to potentially malice code locations.

\item The average FPR and precision value is seemingly poor i.e., 17\% and 14.34\%, respectively. The major reason for high FPR and low precision lies in the ratio between the number of actual and predicted malice classes. The average number of malice classes in the MYST dataset is as low as 2.48. 
However, remember, the malice code in MYST dataset is automatically generated and not very much close to the real-world scenario, where we would have a larger number of malice classes on average. For instance, even in the primitive \textit{ADRD} and \textit{Geinimi} samples that were used in our case studies contained 32 and 82 malice classes, respectively.
Given this statistics, in order to contain FNs, we have considered 10 classes with highest m-scores to be malice indeed.
Since, the ratio of ground truth malice classes and predicted malice classes is very much skewed (i.e., 2.48/10), we obtain low average FPR and precision values.
Furthermore, unlike automated detection, from an analysis view-point, 17\% of FPs is permissible. Meaning, while inspecting roughly 1/7th of the whole code, only 17\% of the times \tool{} would present a false positive class for review to the analyst. Just by reviewing the preliminary details such as name of the class (i.e., whether it is from libraries, or from host app and not the rider code), the analyst could quickly spot these FPs.
Hence, given the nature of our application, low average FPR and precision could be justified.

\end{itemize}
\leavevmode 
\newline
{%group to keep \parindent change local
	\parindent-\fboxsep     %revert indentation due to \fbox frame space
	\indent%    
	\fbox{%
		\parbox{\linewidth}{%
			%			\parindent\defaultparindent%
			\indent \textit{Summarizing RQ3 inferences, it is clear that \tool's individual views, owing to their inherent limitations, could only  reveal a minority of malicious code portions. \tool{} achieves more comprehensive and precise malicious code localization results by appropriately combining the m-scores from individual views.}
		}
	}
}%end parindent group

\section{Related Work}
\label{sec:rw}

Many well-known malware detection techniques have been reviewed in previous sections. We throw light on remaining works and contrast them from \tool{} in this section under three categories: (1) PRG based approaches that use only one set of features, (2) approaches that use multiple feature-sets and (3) approaches that attempted coarse-grained malicious code localization.

\subsection{PRG based Android malware detection}
\label{ss:rw_amd}
Structural features from PRGs (subgraphs, walks, etc.) have been used by a family of approaches reviewed below. 

\textsc{Adagio}'s structural detection and comparison has been already discussed in detail in \S\ref{ss:impl} and \S\ref{sec:rd}. 
DroidMiner \cite{DroidMiner} proposes a two-tiered behavior graph to model malicious program logic into a vector of threat modalities, and then applies classification according to these modalities. 
DroidSIFT \cite{DroidSift} models API-relevant behaviors into weighted CADGs and classifies malware based on a vocabulary of known malicious CADG subgraphs. However, unlike \tool{}, DroidSIFT uses a fixed vocabulary of hand-picked subgraphs to construct feature vector representations. This limits the recall of the model when used over a longer period of time.
Recently, AppContext \cite{AppContext}  proposes differentiating malicious and benign behaviors based on the contexts similar to ours. However, it ends up capturing contextual features from individual nodes without their topological structural information.
MassVet \cite{massvet} statically analyzes apps' UI code to extract a graph that expresses UI states and transitions. Subsequently, it uses a  DiffCom analysis to detect repackaged malware.
MaMaDroid \cite{mama} constructs CGs, models API call sequences as Morkov Chain features and uses Instance-based classifiers (RFs and kNNs) for detection. Both MassVet and MaMaDroid do not capture contextual information. 

On the other hand, a prominent set of works which leverage on PRGs for information-flow analysis include FlowDroid \cite{flow}, IccTA \cite{ICCTA}, Mudflow \cite{Mudflow} and DroidSafe \cite{ds}. These works predominantly use only the data-flow view and target detecting privacy leak attacks and related malware.  

In sum, all the above-mentioned approaches capture PRGs' structural information from only one perspective, unlike \tool. 

\subsection{Multi-perspective approaches}
\label{ss:rw_mp}
Recently, quite a few approaches leveraging on multiple features sets with different modalities have been proposed. However, \tool{} differs from them in two ways: (i) none of them use both context-aware and structural features which complement each other yielding better accuracies, and (ii) the way \tool{} appropriately and systematically integrates multiple views.

Prominent works that employ multiple static and dynamic analysis-based feature-sets are discussed below. \textsc{Drebin}'s methodology which uses as many as 8 feature-sets has been explained and compared in \S \ref{ss:impl} and \S \ref{sec:rd}. 
Sahs and Khan \cite{MLMalDetect} extract a variety of features including tokens from user-defined permissions, standard permissions and CFG signatures and subsequently takes an anomaly detection approach using a One-Class SVM to detect malware.
Peiravian et al. \cite{peir}  take a simpler approach by considering permissions and API calls as features. MAST \cite{mast} uses selected permissions, Intent filters, the existence of native code and zip files, then applies Multiple Correspondence Analysis to perform malware detection. MADAM \cite{madam} uses five feature-sets that includes system calls, critical APIs, user-interaction based features and metadata related features (e.g., rating, number of downloads, etc.). RevealDroid \cite{reveal} uses four feature-sets namely, sensitive APIs, information flows, Intent actions and package-level API informations with Decision Tree classifier to perform detection. 
%Recently, several techniques such as MARVIN \cite{marvin}, StatDynA \cite{statdyna} and StormDroid \cite{stormdroid} adopt a hybrid approach by combining both static and dynamic analysis features and demonstrated achieving better accuracies than solutions leveraging on one of the analysis paradigms.

However, all the above-mentioned approaches just perform an early fusion of their multi-modal features i.e., just concatenate the feature vectors from individual perspectives. This results in obtaining a performance on par with uniform kernel of these features. As demonstrated in \S \ref{ss:acc}, this results in sub-optimal accuracies. Unlike these approaches, thanks to its MKL phase, \tool{} arrives at the most appropriate combination of its perspectives that offers best accuracy and explainability.

To the best of our knowledge, the only other work that uses MKL for multi-view Android malware detection is HADM \cite{hadm}. \tool{} differs from this work in the following aspects: (i) HADM uses primitive features such as frequencies of APIs, advertisement network names etc. along with similar structural features. In other words, HADM does not use robust context-aware features like our approach, (ii) HADM uses a non-linear combination of base-kernels and hence its predictions are inexplainable, rendering it incapable of performing malicious code localization.

\subsection{Malicious code localization}
\label{ss:rw_mcl}
As mentioned earlier, multi-perspective malicious code localization is a unique feature of \tool. However, we note that recently, two approaches have attempted to locate malice portion of PRGs. They are reviewed and contrasted below.

\textbf{DrDroid \cite{drdroid}.} This approach is specifically designed to detect repackaged malware and locate the injected malice code. DrDroid achieves this by splitting an app's CG into multiple regions called as DRegions and predicts whether each of them is malicious or benign. However, on many occasions, this method finds only one DRegion in apps, thus labeling all the code in an app as either malicious or benign. In fact, out 5,600 apps in \textsc{Drebin} \cite{Drebin} dataset, this approach marked the entire code as malicious in 3,757 apps. Also, this approach could not rank portions of apps such as methods or classes based on their severity or degree of maliciousness. Meaning, all the code portions in the malice DRegion are considered equally malice.

\textbf{HookRanker \cite{hookranker}.} This approach is designed to identify piggybacked packages that are potentially malice from repackaged version of benign apps. HookRanker makes a strong assumption that malice code in repackaged malware will be contained in separate packages. While the validity of this assumption is debatable, this approach is incapable of identifying finer malice portions like methods and classes. As pointed out in \cite{massvet} and \cite{Genome}, in many cases only certain classes and methods of the injected code are malice.
In fact, HookRanker states \textit{"we consider all
the injected code as malicious, even if the actual malicious payload is only some part of the added code"}.
 %Moreover, HookRanker does not perform semantic malware detection but rather uses names of known malice packages as features, as it only targets finding variants of known malware.
HookRanker is also incapable of discriminating code portions based on their degree of maliciousness.

In-principle, both these approaches are not designed to detect malware that are not piggybacked and hence could not be deployed as general malware detection solutions.

In sum, none of the above-mentioned techniques exhibit all the three qualities that \tool{} possesses: context-aware, multi-view malware detection and  malicious code localization.

\section{Limitations}
\label{sec:lim}

\textbf{Lack of data-flow and dynamic analyses.} The goal of \tool{} is to take context-aware multi-view approach towards  comprehensive malware detection and our evaluations in RQ1 (\S\ref{ss:acc}) demonstrated \tool's efficacy in such a detection. However, \tool, cannot generally detect all sorts of malicious behaviors, as it builds on concepts of static analysis and lacks dynamic inspection. Moreover, information leak attacks identified by our approach are prone to false positive, as it takes into account only control-flow features and lacks data-flow analysis. As a natural extension of \tool{}, we intend to integrate these two perspectives in our future work.
The reason behind not including them in the current work, is their poor scalability\footnote{As discussed in \cite{Drebin,Mudflow,reveal,DroidMiner,androidsurvey} performing precise data-flow and dynamic analysis to extract features is computationally heavy. }. 

In the current work, we mitigate the absence dynamic analysis features by maneuvering our multi-view analysis as follows:
{\tt Permission} view extracts API calls related to obfuscation, reflection, and loading of code, such as {\tt reflect.Constructor} and {\tt DexClassLoader.loadClass} and considers their invocations as use of special permissions. These CPDG features enable us to at least spot the execution of hidden code—even if we cannot further analyze it. In combinations with features from other views, \tool{} is still able to identify malicious behaviors despite the use of some obfuscation techniques.

\textbf{Population drift.}
Recently, malware population drift (i.e., drift in malicious characteristics induced by malware evolution over time) has been studied closely by the research community and considered as serious and legitimate threat to practicality of malware detection techniques \cite{droidol,prescience,condrift}. These works suggest using incremental ML techniques (e.g., online and active learning) to handle this drift, automatically. As noted through our evaluations in RQ 1.4 (\S\ref{sss:rq1.4})
\tool, being a batch-learning based framework, falls short of handling this drift and its recall keeps reducing over time. We intend to address this using Online MKL \cite{omkl} approaches in the future.

\textbf{Adversarial attacks.} Another limitation which follows from the use of ML is the possibility of attacks by adversaries such as poisoning (see \cite{poison,adverdrift}). While common obfuscation strategies, such as identifier renaming and code reordering do not affect \tool, adversaries may succeed in reducing its accuracy by incorporating benign contextual subgraph features or fake invariants into malicious apps.
Even though such adversarial attacks against ML based detectors cannot be ruled out in general, meticulous sanitization of training data (see \cite{adverdrift}) can limit their impact.

%\vspace{-5mm}
\section{Conclusions \& Future Works}
\label{sec:conc}

{\color{black}
In this paper, we propose \tool, a framework that performs context-aware, multi-view malware detection and malicious code localization. In its pipeline of detection and localization process, firstly, \tool{} deploys static analysis and graph kernels to extract five complementary sets of semantic views of apps. Subsequently, it combines these views in a systematic and scalable manner using MKL and performs malware detection. Finally, ensuing detection, \tool{} uses a novel kernel methods based approach to award m-scores to every basic block, method and class in an app which quantifies the degree of malice activities it performs. This helps to precisely locate malice code portions. Through our large-scale experiments on both benchmark and wild dataset apps, we demonstrate that \tool{} outperforms \soa{} techniques in terms of accuracy (particularly, by more than 11\% F-measure on real-world experimental settings), while maintaining high efficiency. Also, in malicious code localization experiments, it identifies all the malice classes in piggybacked malware apps with 94\% average recall.}

\textbf{Future work.} In our future work, we plan to investigate replacing \tool's batch MKL with online MKL algorithms \cite{omkl} so that it automatically adapts to malware evolution and population drift. Another straight-forward extension of our framework would be to use more semantic views towards performing more comprehensive detection. To this end, we plan to incorporate dynamic analysis based features.

\textbf{Release of results.} In order to provide scope for persuasive research on malicious code localization, we release the results of qualitative and quantitative evaluations of all the apps in \textsc{Drebin} \cite{Drebin} and \textsc{Mystique} \cite{mist} datasets at \cite{oursite}. %This dataset contains the ICFGs of all the apps. Each node is assigned an m-score quantifying its degree of maliciousness. 
%Additionally, for every node, sensitive APIs accessed, permissions, sources, sinks used, instruction sequences and CFG signatures are also provided as node attributes. This helps to gain better insights on code portions, similar to the ones presented in our case studies (see \ref{ss:lme}). ICFGs (with aforementioned node attributes) of 1000 MYST apps used in our quantitative evaluation in \S \ref{sss:lme_quant} are also released.

%\appendices

\section{Appendix}
\subsection {Contextual relabeling Algorithm}
\label{app:cr}

\begin{algorithm}[ht]
	\scriptsize
	\caption{CWLK - Contextual relabeling}
	\label{algo:cr}
	\SetKwInOut{Input}{input}
	\SetKwInOut{Output}{output}
	\SetKwInOut{Given}{given}
	\Input
	{$G = G_0 = (N, E, \lambda, \xi)$ | \textit {PRG} with set of nodes ($N$), set of edges ($E$) and set of node labels ($\lambda$) and contexts ($\xi$)\newline
		$h$ | number of iterations 
	}
	\Output
	{
		$\{\mathcal{G}_0, \mathcal{G}_1,...,\mathcal{G}_h\}$ | contextual WL sequence of height $h$
	}
	
	\Begin{
		\For {$i = $ 0 to $h$} 
		{
			\For {$n \in N$}
			{
				$\sigma_i(n) = \{\}$\\
				\If {$i > 0$}
				{
					\tcp{neighbourhood (i.e., degree 1 neighbors) of \textit{n}}
					$\mathcal{N}(n) = \{m\ |\ (n,m) \in E\}$ \\
					$M_i(n) = \{\lambda_{i-1}(m)\ |\ m \in \mathcal{N}(n) \}$\\
					\tcp{neighbourhood label}
					$\lambda_i(n) =  \lambda_{i-1}(n) \oplus sort(M_i(n))$
					
				}	
				\tcp{adding context to the neighborhood label}
				\For {$c \in \xi(n)$}
				{
					$\sigma_i(n) = \sigma_i(n) \cup c \oplus \lambda_i(n) $ 
				}
				\tcp{contextual neighbourhood label}
				$\sigma_i(n) = join(\sigma_i(n))$ \\
				\tcp{optional step: label compression}
				$\gamma_i(n) = f_c(\sigma_i(n))$ 
				
			}
			\tcp{Contextual WL graph at height \textit{i}}
			$\mathcal{G}_i = (N,E,\gamma_i)$ 
		}
		\textbf{return} \{$\mathcal{G}_0,\mathcal{G}_1,...,\mathcal{G}_h$\}
	}
\end{algorithm}

Algorithm \ref{algo:cr} presents the contextual relabeling process. The inputs to the algorithm are PRG, $G$ and the degree of neighborhoods to be considered for re-labeling, $h$. The output is the sequence of CWL graphs, $\{\mathcal{G}_0, \mathcal{G}_1,...,\mathcal{G}_h\}\!=\!\{(N,E,\gamma_0), (N,E,\gamma_1),...,(N,E,\gamma_h)\}$, where $\gamma_1,...,\gamma_h$ are constructed using the contextual relabeling procedure.

For the initial iteration $ i=0 $, no neighborhood information needs to be considered. Hence the contextual neighborhood label $ \gamma_0 (n) $ for all nodes $ n \in N $ is obtained by justing prefixing the contexts to the original node labels to arrive at $\sigma_0(n)$ (lines 9-11). $\sigma_0(n)$ could be optionally compressed with a compression function $f_c$ to compute $\gamma_0(n)$ (line 12).  For $ i\!>\!0 $, the following procedure is used for contextual re-labeling.
Firstly, for a node $n \in N$, all of its neighboring nodes are obtained and stored in $\mathcal{N}(n)$ (line 6). For each node $m \in \mathcal{N}(n)$ the neighborhood label up to degree $i-1$ is obtained and stored in multiset $M_i(n)$ (line 7). $\lambda_{i-1}(n)$, neighborhood label of $n$ till degree $i\!-\!1$ is concatenated to the sorted value of $M_i(n)$ to obtain the current neighborhood label,  $\lambda_i(n)$ (line 8). Finally the current neighborhood label is prefixed with the contexts of node $n$ to obtain $ \sigma_i(n) $ which is then compressed to arrive at, $\gamma_i(n)$, the contextual neighborhood label (lines 9-12). For every iteration \textit{i}, this process of contextual relabeling yields CWL graph at height \textit{i}, $\mathcal{G}_i$ (line 13). Finally, the CWL sequence comprising CWL graphs from height $0$ to $h$ are returned (line 14).

\subsection {Qualitative Analysis of Base Kernels}
\label{app:qual}

\begin{figure*}[t]
	\captionsetup{justification=centering}
	\includegraphics[height=8.5cm,width=15cm]{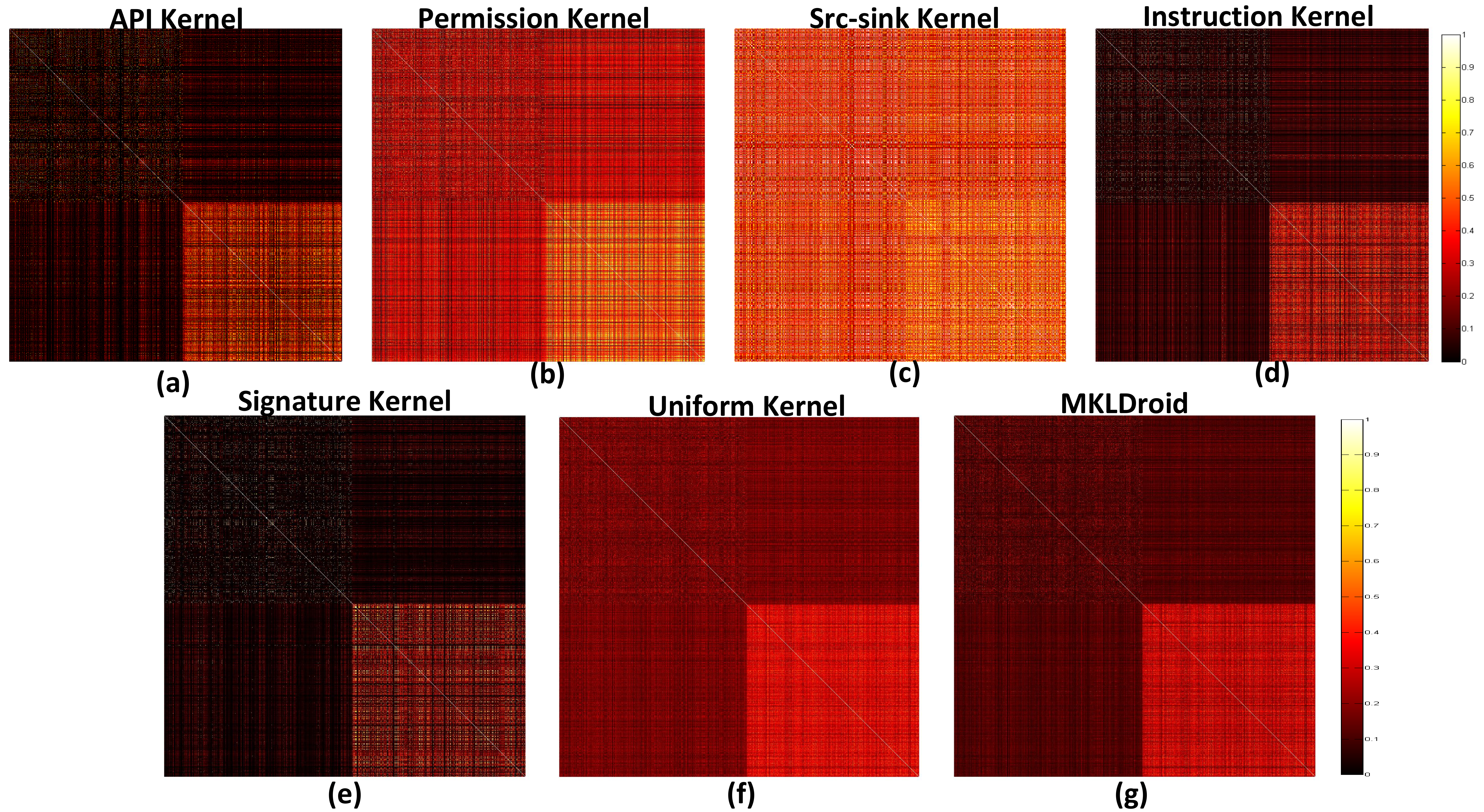}
	\caption{Qualitative Analysis: Comparison of base kernels and kernel combinations through visualizing kernel matrices as heatmaps \label{fig:heatmaps}}
\end{figure*}

The detection capabilities of the base kernels and kernel combinations could also be inferred by visualizing the kernel matrices. To this end, we present the kernel matrix of all the samples used in {\tt CE1} as a heat map in Fig. \ref{fig:heatmaps}. The first (top-left) 5,000 samples are the benign apps from the GP1 collection and the subsequent (bottom-right) 5,600 samples malware from DR collection. Every cell in the kernel matrix represents the similarity value between a pair of apps.
Dark and light shades in cells indicate low and high similarity values, respectively.

It could be clearly seen that the malware apps exhibit high similarities among them in all the views compared to the benign apps. This qualitative depiction reinforces the observations on homogeneity in DR collection that we  discussed above. Also, the inferences on individual base kernel's detection potentials discussed in RQ1.1 are observed qualitatively from figures \ref{fig:heatmaps} (a) - (e). For instance, the {\tt API} kernels separates the malware and benign samples better than other base kernels. Also, the non-uniform linear combination of kernels learnt by SMO-MKL (Fig. \ref{fig:heatmaps} (g)) offers the best separation between the samples of the two classes.

\end{document}